\documentclass{PoS}

\usepackage{cancel}

\newcommand{\nc}[1]{\newcommand{#1}}

\nc{\its}[1]{\itshape #1 \upshape}
\nc{\mc}[3]{\multicolumn{#1}{#2}{#3}}

\nc{\bc}{\begin{center}}
\nc{\ec}{\end{center}}
\nc{\ig}[1]{\bc \includegraphics{#1} \ec}

\newcommand{\hmu}{\hat{\mu}}

\nc{\bo}[1]{\mbox{\boldmath \( #1 \! \! \)  \unboldmath}}
\nc{\nnn}{\nonumber \\}
\nc{\f}[2]{\frac{#1}{#2}}
\nc{\td}[2]{\f{d #1}{d #2}}
\nc{\pd}[2]{\f{\partial #1}{\partial #2}}
\nc{\suli}{\sum\limits}
\nc{\proli}{\prod\limits}
\nc{\ili}{\int\limits}
\nc{\sr}[2]{\stackrel{#1}{#2}}
\nc{\dps}{\displaystyle}
\nc{\ket}[1]{\left| #1 \right>}
\nc{\bra}[1]{\left< #1 \right|}
\nc{\bracket}[2]{\left< #1 \right| \left. \! #2 \right>}
\nc{\norm}[1]{\left\| #1 \right\|}
\nc{\lndm}[1]{\pd{^{#1} \ln{\det{M}}}{\mu^{#1}}}
\nc{\pdmm}[1]{M^{-1} \pd{^{#1} M}{\mu^{#1}}}
\nc{\pdm}{M^{-1}\pd{M}{\mu}}
\nc{\trac}[1]{\mbox{Tr}\left(#1\right)}
\nc{\hm}{\hat{m}}

\newcommand{\be}{\begin{equation}}
\newcommand{\ee}{\end{equation}}
\newcommand{\bea}{\begin{eqnarray}}
\newcommand{\eea}{\end{eqnarray}}
\newcommand{\bean}{\begin{eqnarray*}}
\newcommand{\eean}{\end{eqnarray*}}

\title{Lattice QCD at nonzero temperature and density}

\ShortTitle{LQCD at nonzero temperature and density}

\author{\speaker{Heng-Tong Ding}\thanks{The work is supported by the National Natural Science Foundation of China under the grant no. 11535012.}\\
        Key Laboratory of Quark \& Lepton Physics (MOE) and Institute of
Particle Physics, \\
Central China Normal University, Wuhan 430079, China\\
        E-mail: \email{hengtong.ding@mail.ccnu.edu.cn}}

\abstract{In this talk I review the current status of lattice QCD calculations at nonzero temperature and density. I focus on the QCD phase structure and bulk QCD thermodynamics at zero and nonzero chemical potentials.}

\FullConference{34th annual International Symposium on Lattice Field Theory\\
		24-30 July 2016\\
		University of Southampton, UK}

\begin{document}

\section{Introduction}

Lattice QCD has been an excellent tool to study the properties of strong-interaction matter under extreme conditions, e.g. high temperature $T$, nonzero chemical potentials $\mu$ and
strong magnetic field etc. Two long-time goals of lattice QCD computations at nonzero temperature have been completed in the past years: one is Equation of State at vanishing baryon density in a wide temperature range and the other is the nature of the transition from hadronic phase to quark gluon plasma phase. With lattice QCD many more knowledge on the thermodynamic as well as the dynamic properties of the QCD medium under extreme conditions have been acquired. 

Many remarkable achievements have been made since the lattice QCD conference in the year of 2015. Due to the limitation of the length of the proceedings, I restrict myself reviewing several selected topics at $T>0$ and $\mu=0$ in Sections~\ref{sec:Phase} and~\ref{sec:Thermodynamics}, and at $T>0$ and $\mu>0$ in Sections~\ref{sec:EoSMu} and Sections~\ref{sec:PhaseMu}. 
Here Section~\ref{sec:Phase} includes topics on the QCD phase structure in the quark mass plane, Section~\ref{sec:Thermodynamics} is on the QCD equation of state and topological susceptibility at high temperature,
Section~\ref{sec:EoSMu} on the QCD equation of state at nonzero baryon number density, and Section~\ref{sec:PhaseMu} on the QCD phase structure at nonzero chemical potentials.

 For topics that are not covered and exhaustively discussed here please find them in recent reviews~\cite{Ding:2015ona,Meyer:2011gj,Schmidt:2017bjt}, proceedings of  recent lattice plenary talks~\cite{Sexty:2014dxa,Meyer:2015wax,Borsanyi:2015axp,Bazavov:2015qsa,DElia:2015rwa,Scorzato:2015qts} as well as lattice conference write-ups.

\section{QCD phase structure in the quark mass plane}
\label{sec:Phase}

The nature of the QCD transition depends crucially on the values of the quark masses $m_q$ and
the number of flavors ($N_f$).  Basic features of this transition can be
understood by invoking universality arguments \cite{Pisarski:1983ms}. 
These features are summarized as follows:
 {\bf 1)} in the limit of $m_q\rightarrow 0$ or $m_q\rightarrow \infty$ for $N_f=3$ QCD, there exists a first order phase transition. The first order phase transition regions
end at two second order phase transition lines belonging to the universality class of the 3-d, Z(2) symmetric  
Ising model.  These two second order phase transition lines  are separated by a cross over region; 
{\bf 2)} In the limit of $m_q\rightarrow0$ for $N_f=2$ QCD there are two cases: a)  if $U_A(1)$ symmetry is broken there exists a second order phase transition belonging to O(4) universality class; b) if $U_A(1)$ symmetry is effectively restored
the transition will be of first order~\cite{Pisarski:1983ms} or a second order belonging to O(2)$\times$O(4) universality class\cite{Pelissetto:2013hqa,Grahl:2013pba} ;
{\bf 3)} If the QCD phase transition is of second order in the limit of $m_q\rightarrow0$  there exists a tri-critical point where three different types of transitions meet.

The above features of QCD phase structure can only be confirmed through non-perturbative calculations.
Recent investigations on the above issues from lattice QCD computations are shown in the following subsections.

\subsection{Boundary of the first order chiral/deconfinement phase transition region}
\label{sec:boundary}

The first order chiral phase transition in 3-degenerate flavor QCD has been investigated on 
coarse lattices using unimproved~\cite{Christ:2003jk, deForcrand:2003ut,deForcrand:2007rq,Smith:2011pm} as well as improved actions
~\cite{Karsch:2001nf,Schmidt:2002uk,Karsch:2003va, Bernard:2004je,Cheng:2006aj,Endrodi:2007gc,Jin:2014hea,Varnhorst:2015lea}.
However, the value of pion mass at the critical point where the first order phase transition ends, $m_\pi^c$, is 
not yet determined in the continuum limit and its value at finite lattice cutoff varies from $\sim$300 MeV to $\sim$70 MeV, 
strongly depending on the lattice spacing and discretization schemes of the action.
By using the standard staggered action it was found that the critical pion mass $m_\pi^c$ is about 300 MeV\cite{Christ:2003jk, deForcrand:2003ut} while  
using the improved p4 action $m_\pi^c$ is about 70 MeV~\cite{Karsch:2003va}. These results are obtained on the lattices with the 
temporal extent $N_{\tau}=4$. Other calculations using p4 and asqtad actions on finer lattices disfavor a critical pion mass of 300 MeV~\cite{Cheng:2006aj,Bernard:2004je}.
On finer lattices with $N_{\tau}=6$ it has been shown that $m_\pi^c$ becomes smaller compared 
with the one obtained from simulations on $N_{\tau}=4$ lattices using standard staggered fermions~\cite{deForcrand:2007rq}.  

\begin{figure}[htp]
\begin{center}
\includegraphics[width=0.45\textwidth]{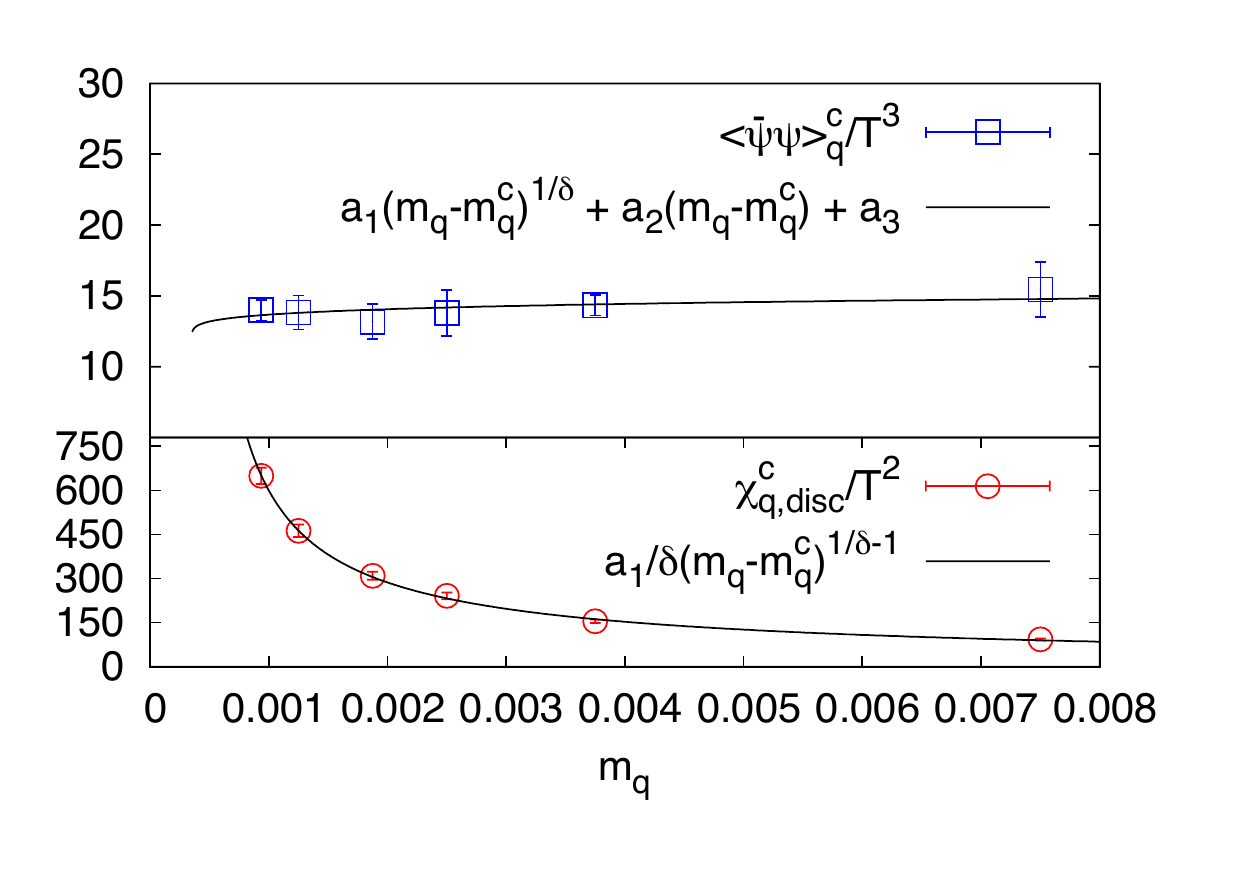}~\includegraphics[width=0.45\textwidth]{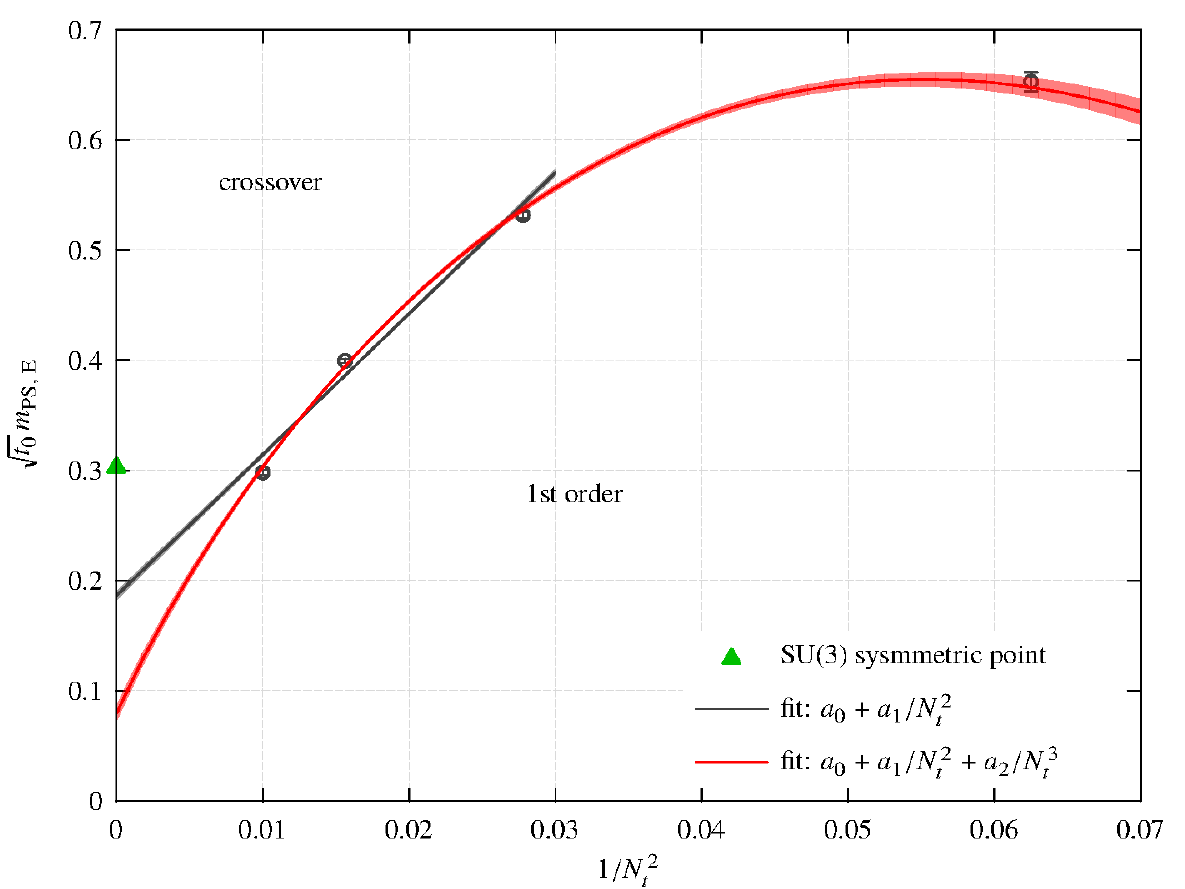}
\end{center}
\caption{State-of-the-art estimates on the critical pion mass where the first order chiral phase transition ends on the second order Z(2) transition line. Left: Simulations carried out using HISQ fermions on $N_\tau=6$ lattices~\cite{Bazavov:2017xul}. An upper bound of critical pion
mass is obtained, i.e. $m_\pi\lesssim50$ MeV. 
Right: Preliminary results on the continuum extrapolation of $m_\pi^c$ obtained from simulations using clover-improved Wilson fermions on $N_\tau=4$, 6, 8 and 10 lattices~\cite{Takeda:2016vfj}. The critical pion mass in the continuum limit
is estimated to be around 100 MeV.}
\label{fig:mc3fQCD}
\end{figure}

Recently the first order chiral phase transition region has been investigated using Highly Improved Staggered Quark action on $N_\tau=6$ lattices~\cite{Bazavov:2017xul}.
It is expected that the chiral order parameter $M$ at the pseudo critical temperature  should have a scaling behavior of $M\sim(m-m_c)^{1/\delta}$ as the system is close to the Z(2) critical line. Here $m$ is the quark mass and $m_c$
is the critical quark mass. Though the chiral condensate is not the true order parameter for the chiral phase transition in the 3-flavor QCD it is argued that $\langle\bar{\psi}\psi\rangle\sim(m-m_c)^{1/\delta}$ 
holds in the case of small $m$. By performing a simultaneous
scaling fit to both chiral condensates and disconnected chiral susceptibilities as shown in the left panel of Fig.~\ref{fig:mc3fQCD} an upper bound for the critical pion mass is obtained $m_\pi^c\lesssim 50$MeV~\cite{Bazavov:2017xul}.

Most recently the study has also been performed using improved clover-Wilson fermions~\cite{Jin:2014hea} on $N_\tau=$4, 6, and 8 lattices. The estimate critical pion mass in the continuum limit is 
reduced from $\sim 300$ MeV to $\sim$ 100 MeV after including the analyses on the new data produced on $N_\tau=$10 lattices~\cite{Takeda:2016vfj}. Uncertainties arising from the Binder cumulant of the chiral condensate rather than that of a proper order parameter are also discussed in S. Takeda's talk~\cite{Takeda:2016vfj}.

Concerned with the rooting issues in the staggered theory P. de Forcrand presented in this conference his study on QCD with 4-degenerate flavor
using unimproved staggered fermions. The simulations have been performed on lattices with $N_\tau=4$, 6, 8 and 10. The general conclusion is that the critical quark mass 
becomes smaller as approaching to the continuum limit~\cite{Forcrand2016}.

All these studies mentioned above suggested that the chiral first order phase transition could be small and this region may has mild influence to thermodynamics happening in the real world. Alternatively the 2nd order phase transition 
belonging to the O(4) universality class, which will be discussed in the next subsection, would be more relevant to the thermodynamics at the physical point.

 The value of the critical pion mass where the first order deconfinement phase transition starts to end on the 2nd order Z(2) phase transition line has been studied quite a while ago on $N_\tau=4$ lattices using the standard Wilson action~\cite{Saito:2011fs}. An updated study has been presented for $N_f=2$ QCD using the standard Wilson action on $N_\tau=8$ lattices~\cite{Czaban:2016yae}.
 The estimated $m_\pi^c$ is within the range of 2.9 GeV$< m_\pi^c <$4.7 GeV and note that $am_\pi$ is larger than 1 in the simulation~\cite{Czaban:2016yae}. Generally the requirement of $am_\pi <1$ in the lattice simulation makes the continuum extrapolation costly in the heavy quark mass region.

\subsection{Chiral phase transition in $N_f$=2+1 and $N_f=2$ QCD}
\label{sec:chiral}
\begin{figure}[htp]
\begin{center}
\includegraphics[width=0.4\textwidth]{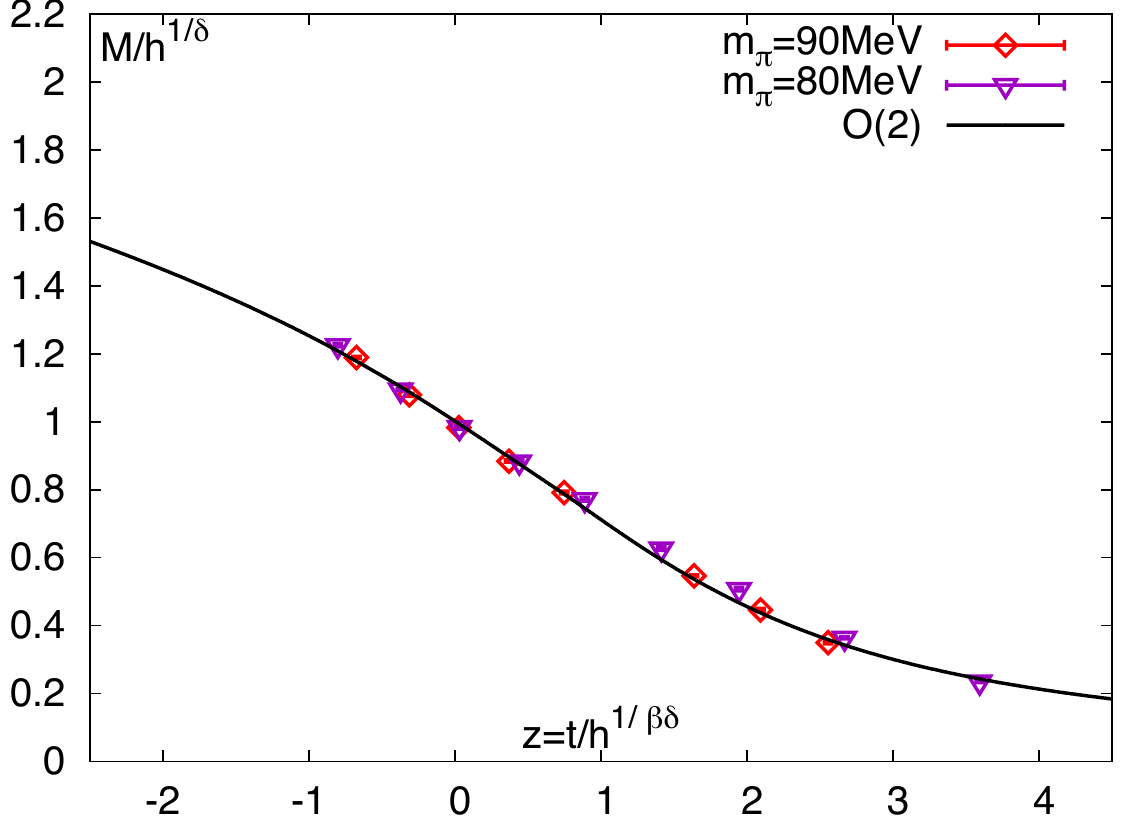}~~~\includegraphics[width=0.35\textwidth]{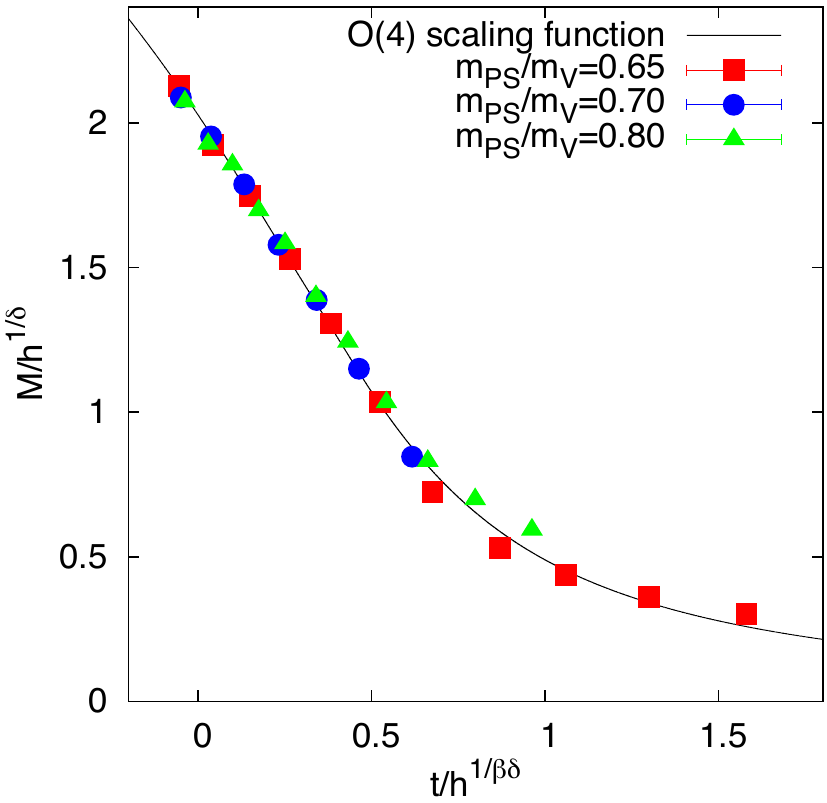}
\end{center}
\caption{Left: O(2) scaling behavior observed based on simulations of $N_f=2+1$ QCD using HISQ fermions on $N_\tau$=6 lattices~\cite{LiLattice2016}. Right: O(4) scaling behavior from simulations of $N_f=2$ QCD using clover-improved Wilson fermions on $N_\tau=4$ lattices~\cite{Umeda:2016qdo}. }
\label{fig:O4}
\end{figure}

In the vicinity to a second order phase transition the order parameter (magnetization) $M$ of system can be described by a single scaling function $f_G$, i.e. Magnetic Equation of State (MEoS)
\be
M/h^{1/\delta}=f_{G}(t/h^{1/\beta\delta}).
\ee
Here $h$ is the external field which is proportional to quark mass, and 
$t=\frac{1}{t_0}\frac{T-T_c(h=0)}{T_c(h=0)}$
is the reduced temperature. They describe the proximity of the system to the critical window. $\beta$ and $\delta$ are critical exponents which are universal for theories with the same global symmetry.
By looking into the temperature and quark mass dependences of M and/or its susceptibilities the chiral phase transition of $N_f=2$ QCD has been studied on the lattice using both Wilson ~\cite{Iwasaki:1996ya,Aoki:1998wg,PhysRevD.63.034502,Burger:2011zc,Bornyakov:2011yb} and staggered fermions~\cite{Karsch:1993tv,Karsch:1994hm,
Laermann:1998gf,Bernard:1996zw,Bernard:1999fv,DElia:2005bv,Mendes:2007ve}. In all these mentioned investigations 
the $U_A(1)$ symmetry is implicitly assumed to be broken,

S.-T. Li from the Bielefeld-BNL-CCNU collaboration presented the updated scaling analyses of chiral observables in $N_f=2+1$ QCD using HISQ fermions on $N_\tau=$6 lattices~\cite{LiLattice2016}.
It is found that in general the scaling violations become larger compared to previous studies using the p4fat3 action. As seen from the left panel of Fig,~\ref{fig:O4} the chiral condensates obtained with $m_\pi=80$ and 90 MeV can be described by the MEoS without an regular term. A Z(2) scaling fit was also performed to the chiral condensate and the critical quark mass $m_c$ is compatible with zero within the errors.
This suggests that $m_s^{tri}<m_s^{phy}$. While in the case that three quark flavors are not degenerate and the ratio of light to strange quark is 
 fixed to be about 1/27 $m_\pi^c$ is suggested to be around $50 $ MeV from the study using the stout fermions on $N_\tau=6$ lattices~\cite{Endrodi:2007gc}.

T. Umeda from the WHOT-QCD Collaboration presented in this conference the updated study on the chiral phase transition in $N_f$=2 QCD~\cite{Umeda:2016qdo}. The simulations were carried out
using clover-improved Wilson fermions on $16^3\times4$ lattices. As seen from the right panel of Fig.~\ref{fig:O4} that the rescaled chiral condensate, which is defined by a Ward-Takahashi identity, falls on the O(4) scaling curve.
This suggests a 2nd order phase transition in the massless two-flavor QCD. Same conclusion was drawn from the so-called many flavor approach~\cite{Yamada:2016hvz,Ejiri:2012rr}, although a first order transition is also suggested from a novel approach using simulations carried out at imaginary chemical potentials using standard staggered fermions~\cite{Bonati:2014kpa} and unimproved Wilson fermions~\cite{Philipsen:2016hkv} on $N_\tau=4$ lattices.

\subsection{Fate of $U(1)_A$ symmetry at nonzero temperature}
\label{sec:axialU1}

As mentioned previously the nature of the chiral phase transition may depend crucially on the strength of $U(1)_A$ symmetry breaking.
The key question is whether $U(1)_A$ symmetry is restored at the same temperature as the $SU_L(2)\times SU_R(2)$ symmetry.
In general two-point correlation functions of pseudo-scalars and scalars are used to probe the restoration of the $U(1)_A$ and $SU_L(2)\times SU_R(2)$ symmetries.
E.g.: the degeneracy between the pseudo-scalar/iso-vector $\pi$ and the scalar/iso-scalar, $\sigma$ (the scalar/iso-vector $\delta$ and the pseudo-scalar/iso-scalar $\eta$) signals
the restoration of the $SU_L(2)\times SU_R(2)$  symmetry, while the degeneracy between $\pi$ and $\delta$ ($\sigma$ and $\eta$) is for the restoration of the $U(1)_A$ symmetry.

\begin{figure}[htp]
\begin{center}
\includegraphics[width=0.25\textwidth]{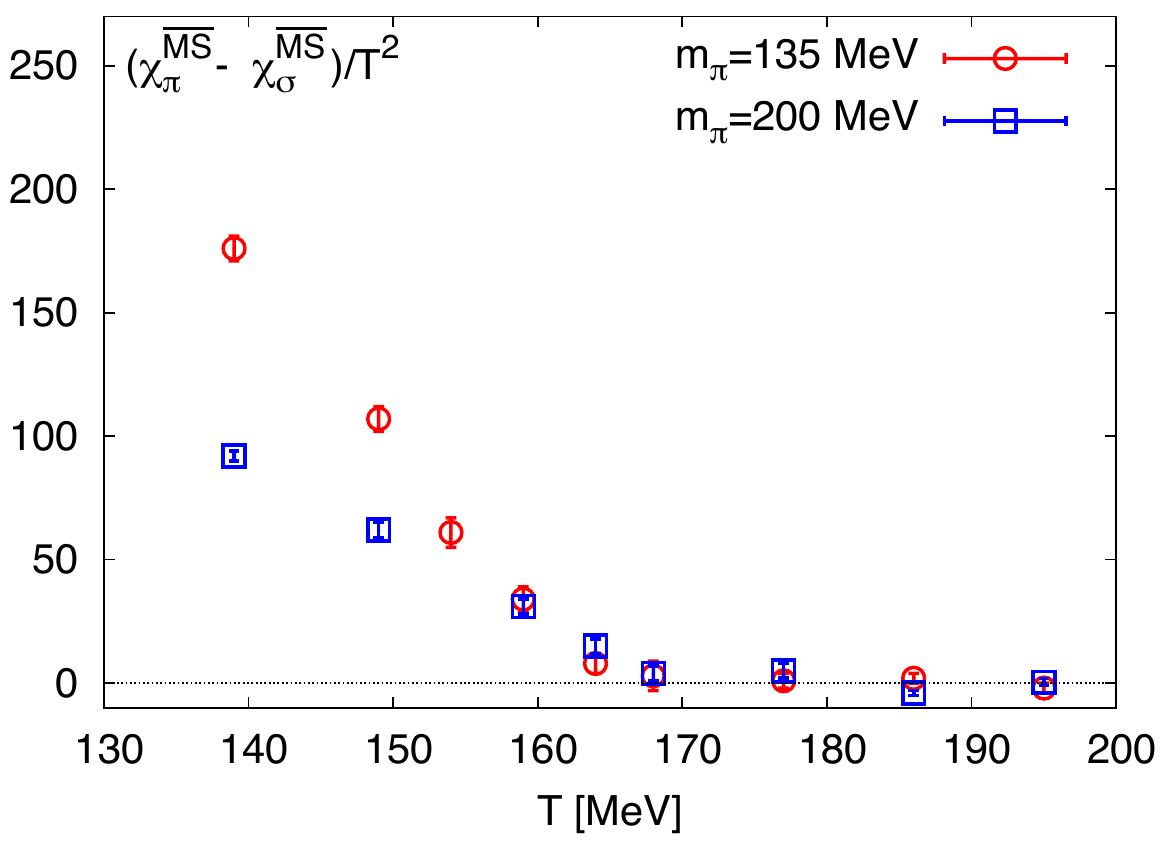}\includegraphics[width=0.25\textwidth]{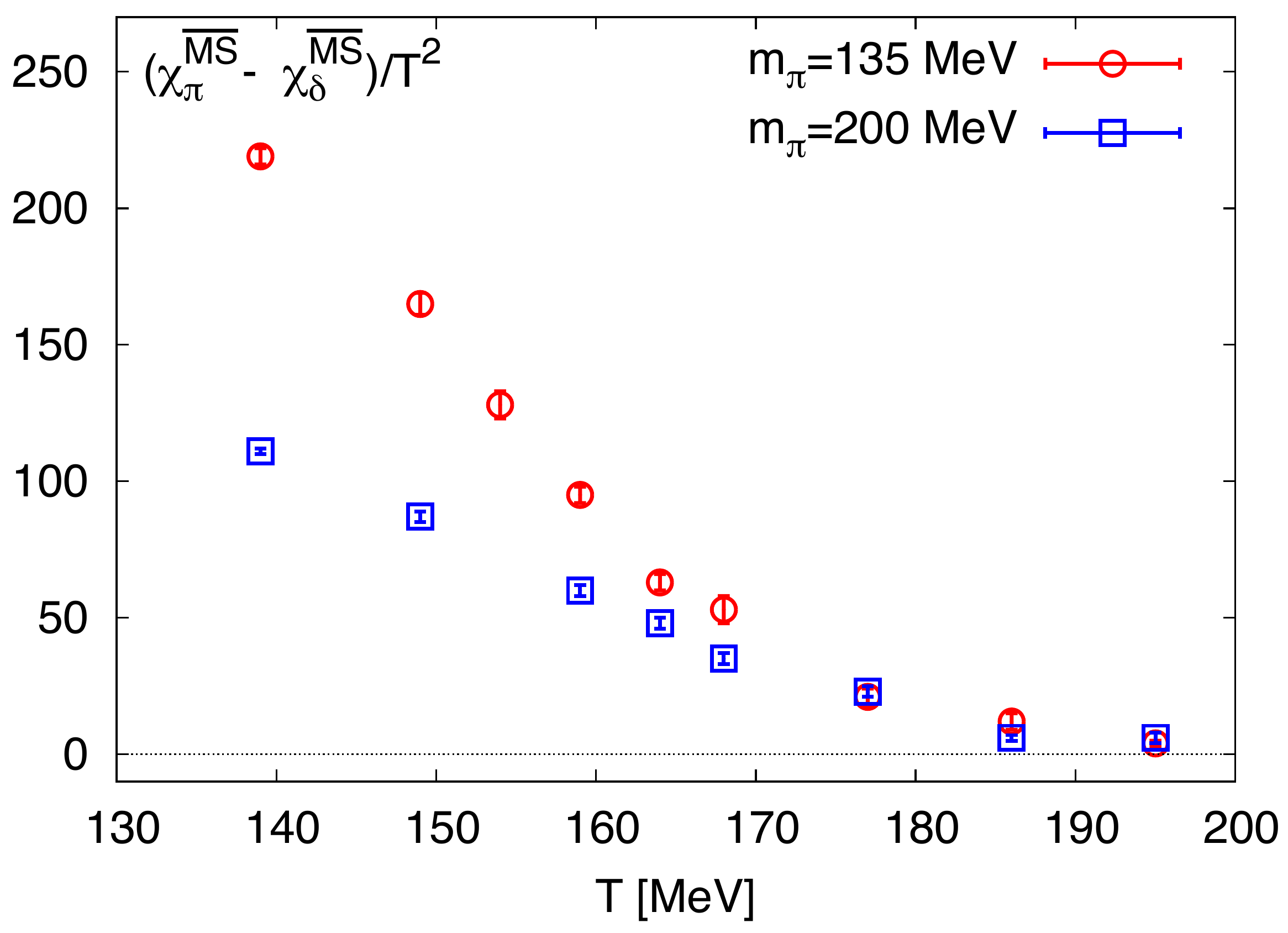}\includegraphics[width=0.25\textwidth]{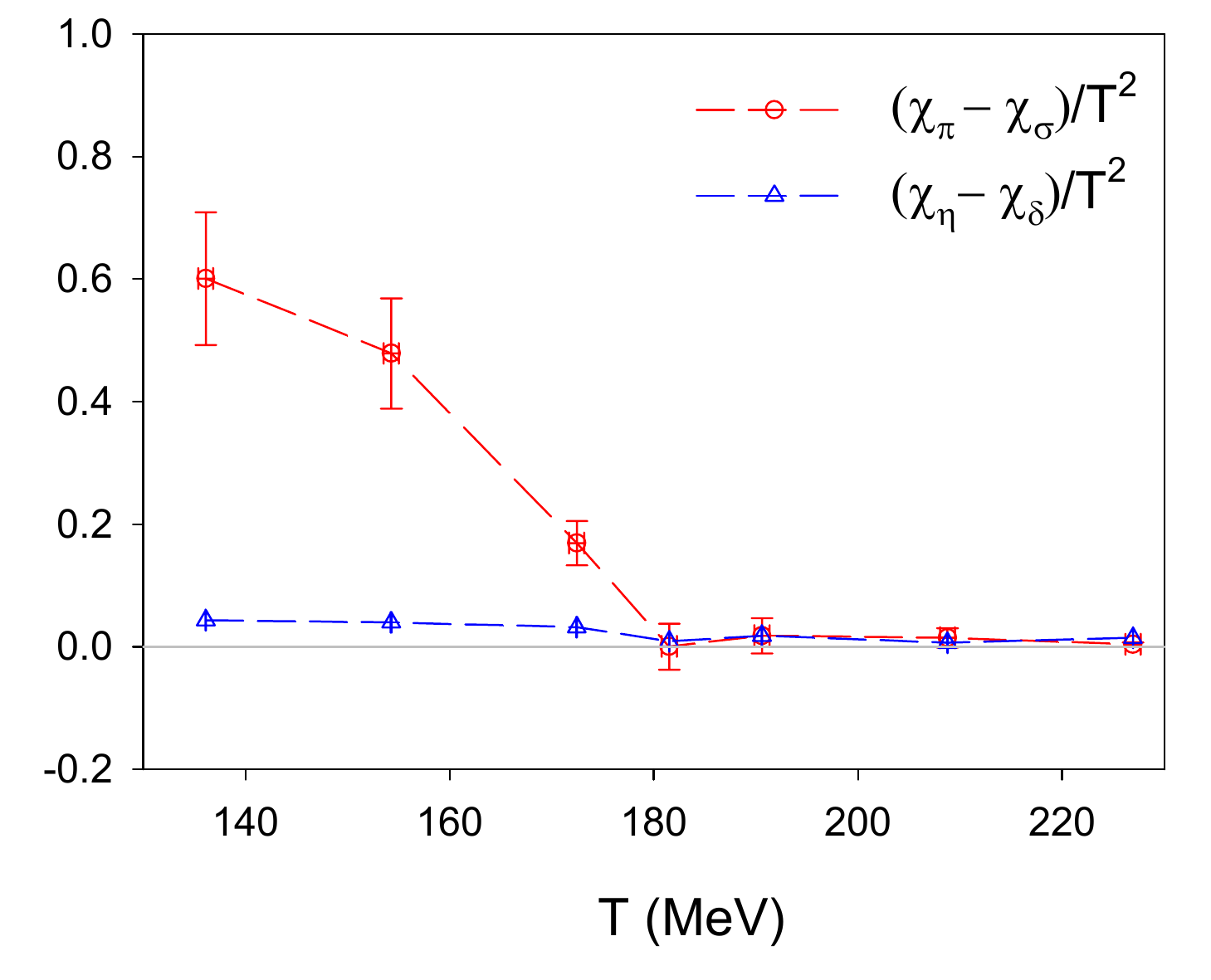}~\includegraphics[width=0.25\textwidth]{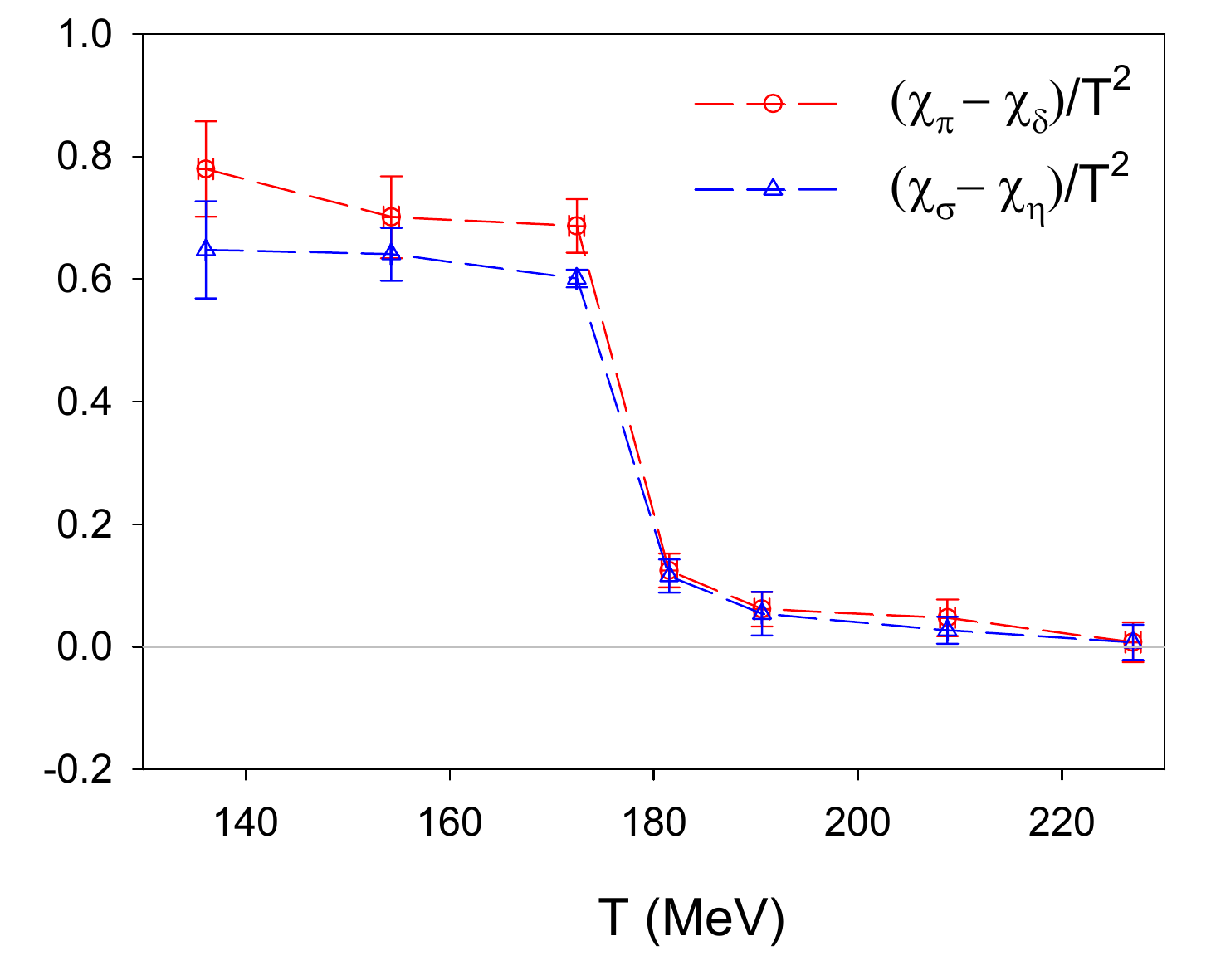}
\end{center}
\caption{Temperature dependences of  $SU_L(2)\times SU_R(2)$ ($\chi_\pi-\chi_\sigma$)  and $U(1)_A$ ($\chi_\pi-\chi_\delta$) susceptibilities. Two plots on the left: Simulations carried out in $N_f=2$ QCD using Domain Wall fermions on $N_t=8$ lattices with $m_\pi=200$ and 135 MeV
~\cite{Bazavov:2012qja,Buchoff:2013nra,Bhattacharya:2014ara}. 
Two plots on the right: Studies performed in $N_f=2$ QCD using the Optimal Domain Wall fermions on $16^3\times 4\times16$ lattices~\cite{Chiu:2013wwa}.}
\label{fig:chi_UA1}
\end{figure}

The study has been carried out in $N_f=2+1$ QCD using M\"obius Domain Wall fermions on $N_\tau=8$ and $N_5$=16-32 lattices with $m_\pi=200$ and 135 MeV~\cite{Bazavov:2012qja,Buchoff:2013nra,Bhattacharya:2014ara}.
Shown in the two left plots of Fig.~\ref{fig:chi_UA1} is $\chi_\pi-\chi_\sigma$ and $\chi_\pi-\chi_\delta$ for two different values of pion masses obtained from these studies.
Here susceptibilities $\chi_{\pi,\delta,\sigma,\eta}$ are the four volume integration of the corresponding two-point correlation functions. It can be clearly seen that at the temperature
when $\chi_\pi-\chi_\sigma$ becomes vanishing, i.e. the $SU_L(2)\times SU(2)_R$ symmetry is restored, $\chi_\pi-\chi_\delta$ remains nonzero and thus the $U(1)_A$ symmetry is still broken. Only at $T\gtrsim 200 $MeV  it starts
to be restored. The right two plots of Fig.~\ref{fig:chi_UA1} show the results obtained in $N_f=2$ QCD using Optimal Domain Wall fermions on $N_\tau=6$ and $N_5$=16 lattices~\cite{Chiu:2013wwa}. Similar conclusions on the fate of the $U(1)_A$ symmetry can be drawn.

\begin{figure}[htp]
\begin{center}
\includegraphics[width=0.45\textwidth]{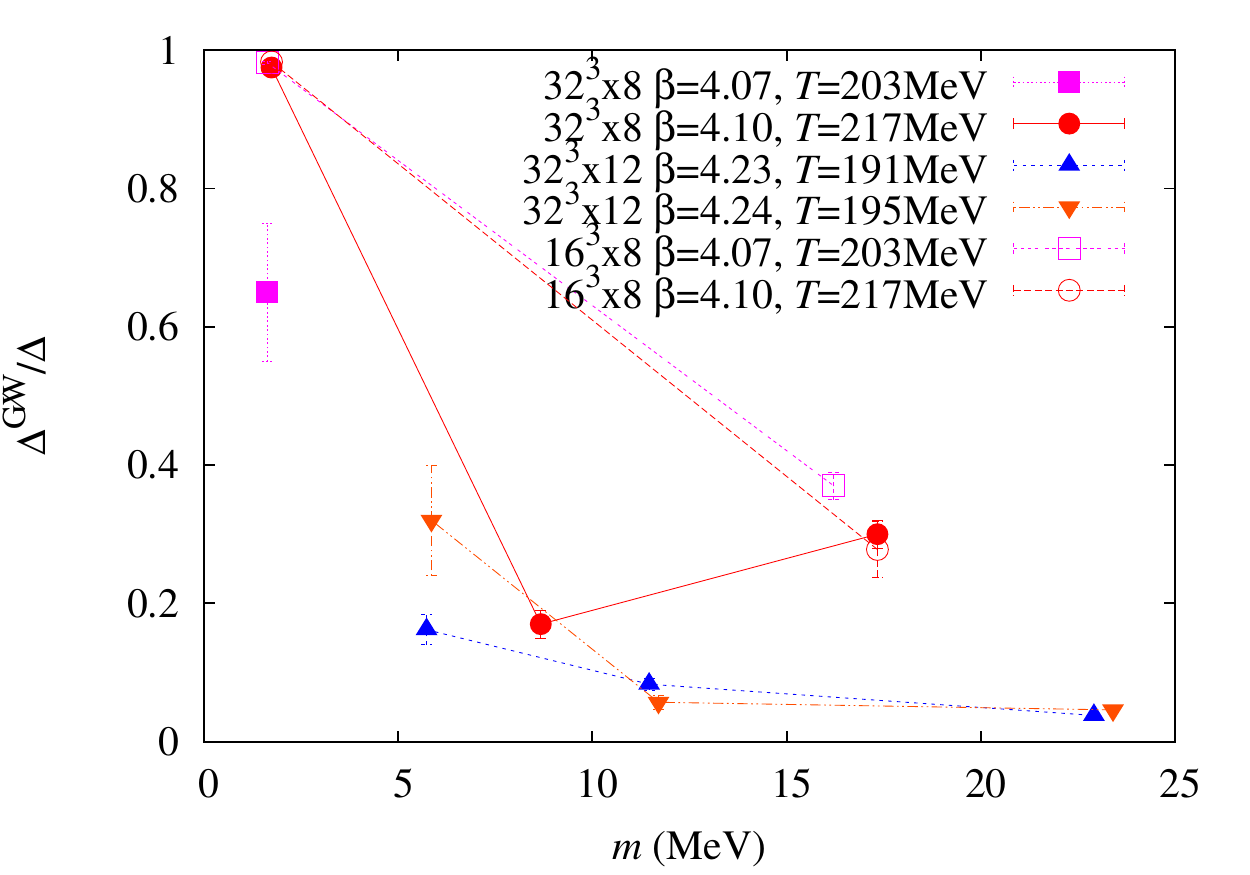}~~~\includegraphics[width=0.45\textwidth]{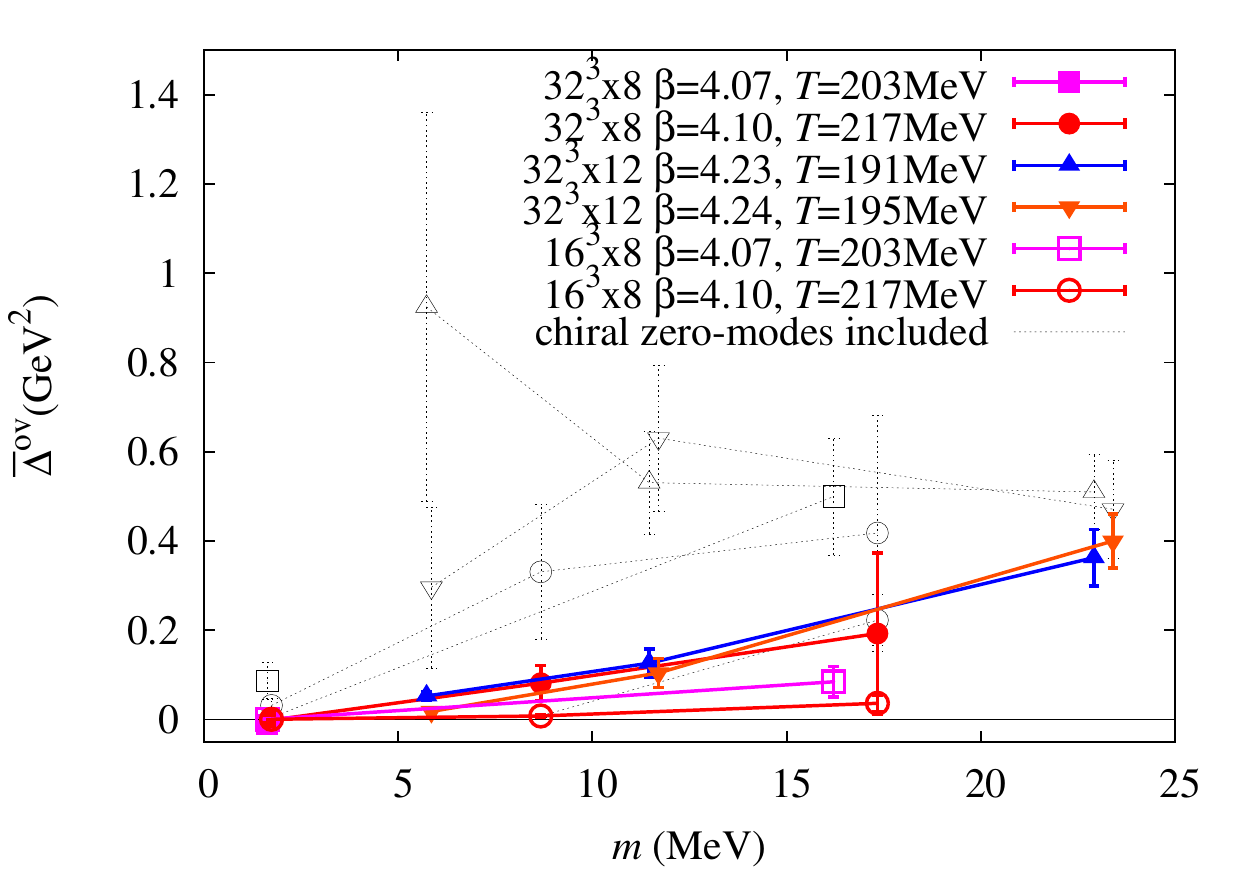}~
\end{center}
\caption{Left: Quark mass and temperature dependences of the fraction of the chiral violation contribution to $\chi_\pi-\chi_\delta$. Right: Quark mass and temperature dependences of $\bar{\Delta}^{ov}$ (solid symbols) and ${\Delta}^{ov}$ (dashed symbols). Figures are taken from Ref.~\cite{Tomiya:2016jwr}. }
\label{fig:JLQCD}
\end{figure}

Very recently the JLQCD collaboration updated their previous studies~\cite{Cossu:2013uua} by carrying out simulations in $N_f=2$ QCD using M\"obius Domain Wall fermions on $N_\tau=8$ and 12 lattices in a temperature window above the critical temperature~\cite{Tomiya:2016jwr}.
By decomposing the $U_A(1)$ susceptibility $\Delta=\chi_\pi-\chi_\delta$ into a chiral symmetric part and a chiral violation part $\Delta^{\cancel{GW}}$ it is found that the chiral violation part dominates in $\Delta$ in particular in the chiral limit as seen from the 
left plot of Fig.~\ref{fig:JLQCD}. Through the reweighting of the Domain Wall fermion determinant to that of the Overlap fermion the $U(1)_A$ susceptibility $\Delta^{OV}$ is obtained and its dependences on quark masses and temperature
are shown as dashed symbols in the right panel of Fig.~\ref{fig:JLQCD}. The  $U(1)_A$ susceptibility with chiral zero modes subtracted, $\bar{\Delta}^{OV}$, is also shown as colored symbols in the plot. It can be seen from the plot that $U(1)_A$ symmetry is 
restored at $T\gtrsim $ 200 MeV in the chiral limit. This is consistent with the findings in Ref.~\cite{Bazavov:2012qja,Buchoff:2013nra,Bhattacharya:2014ara,Chiu:2013wwa}.

\begin{figure}[htp]
\begin{center}
\includegraphics[width=0.33\textwidth]{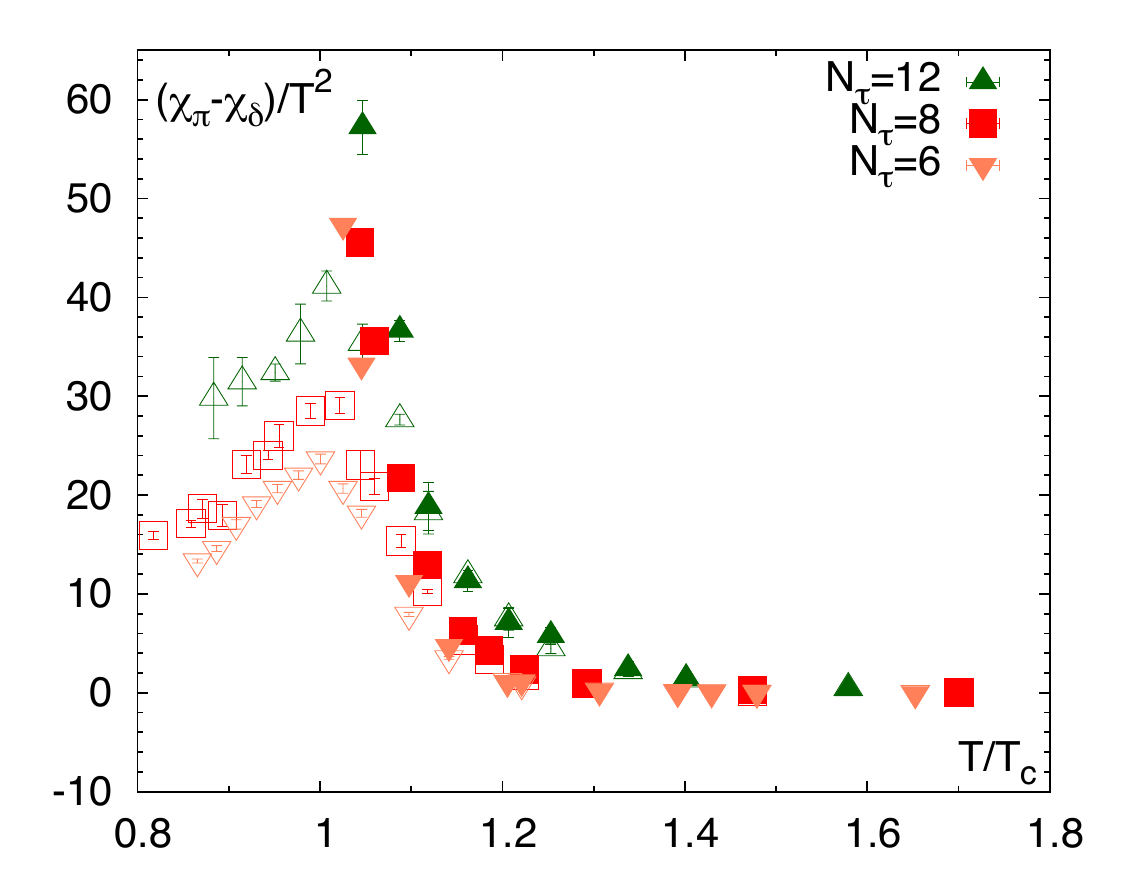}~\includegraphics[width=0.33\textwidth]{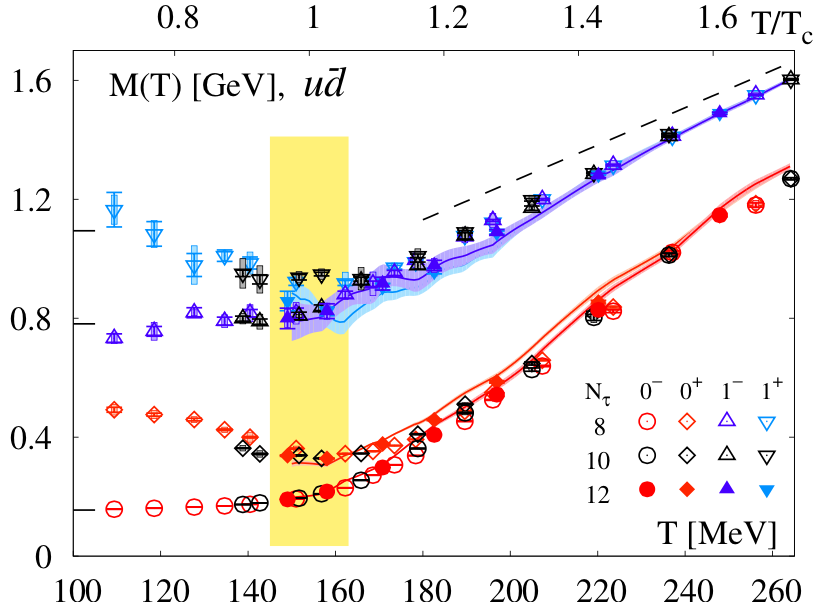}~~~\includegraphics[width=0.33\textwidth,height=4cm]{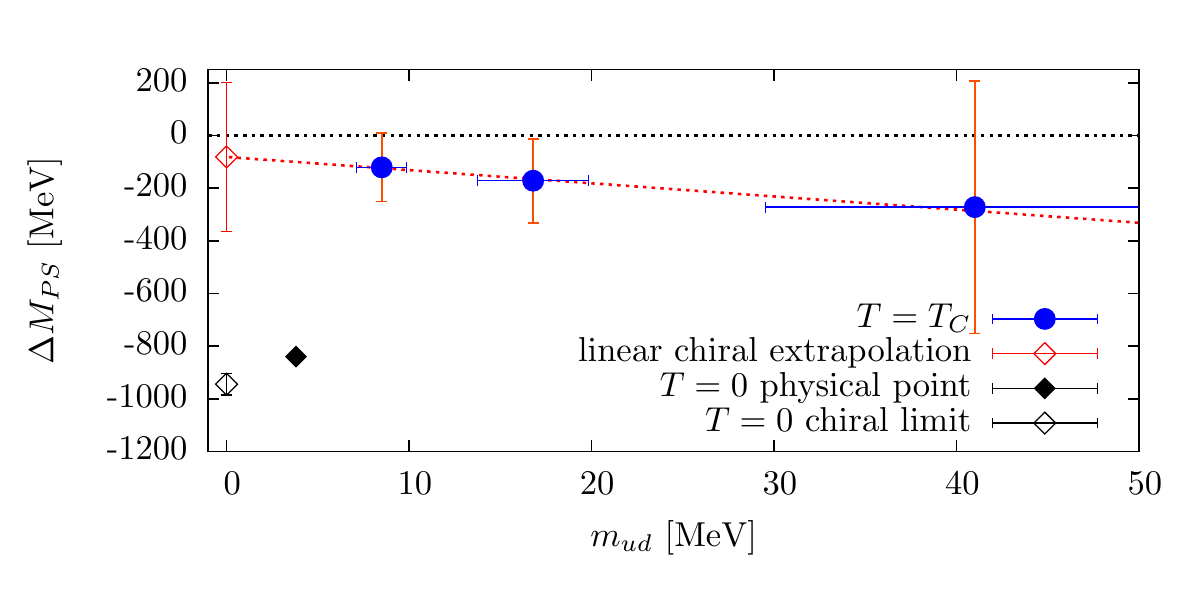}~
\end{center}
\caption{Left:  $\chi_\pi-\chi_\delta$ obtained from simulations of $N_f$=2+1 QCD using HISQ fermions with $m_\pi\approx$160 MeV on $N_\tau$=6, 8, 10 and 12 lattices~\cite{Petreczky:2016vrs}. Middle: Screening mass obtained from two-point correlation functions in various channels of $J^{P}=0^{-}$(pseudo-scalar), $0^{+}$ (scalar), $1^{-}$ (vector) and $1^{+}$ (axial-vector) measured on $N_\tau=8$, 10 and 12 lattices using HISQ fermions with $m_\pi\approx160$ MeV in $N_f$=2+1 QCD~\cite{Maezawa2016}. 
Right: Chiral extrapolation of the difference between screening masses in the pseudo-scalar channel and the scalar channel at the chiral symmetry restoration temperature $T_C$ of $N_f=2$ QCD. Simulations have been carried out using $O(a)$-improved Wilson fermions on $N_\tau$=16 lattices with three pion masses in the range 200 $<m_\pi <$ 540 MeV~\cite{Brandt:2016daq}.}
\label{fig:NonchiralU1}
\end{figure}

Other than chiral fermions the fate of the axial $U(1)$ symmetry has also been studied using staggered~\cite{Cheng:2010fe,Dick:2015twa,Petreczky:2016vrs} as well as Wilson fermions~\cite{Brandt:2016daq}. The most recent results are shown in Fig.~\ref{fig:NonchiralU1}. Based on simulations of $N_f=$2+1 QCD using HISQ fermions with $m_\pi\approx$160 MeV, the results on $\chi_\pi-\chi_\delta$~\cite{Petreczky:2016vrs} and screening masses~\cite{Maezawa2016} in various channels are shown in the left and middle panels, respectively.
The $U(1)_A$ susceptibility shown in the left panel suggests that the axial U(1) symmetry starts to get restored at $T\gtrsim 1.4$~$T_{pc}$. From the perspective of screening masses 
the degeneracy between pseudo-scalar and scalar channels only shows up till $\sim1.2~T_c$, while as expected the screening masses in the vector and axial-vector  channels become degenerate already in the chiral cross over temperature $T_c$ region 
(shown as yellow band in the middle panel).  On the other hand, the differences of screening masses between pseudo-scalar and scalar channels, $\Delta M_{PS}$, obtained from lattice computations using $O(a)$-improved Wilson fermions with 200 MeV$<m_\pi <$ 540 MeV, does not vanish at the chiral symmetry restoration temperature~\cite{Brandt:2016daq}.

As a short summary, while the fate of the $U(1)_A$ symmetry in the chiral limit remains elusive as much more needs to be understood, e.g. detailed temperature and quark mass dependences of $\chi_\pi-\chi_\delta$ as well as $\Delta M_{PS}$, the $U(1)_A$ symmetry is clearly broken at physical pion masses from current studies.

\subsection{Chiral and deconfinement aspects of the QCD transition }
\label{sec:connection}
\begin{figure}[htp]
\begin{center}

~~\includegraphics[width=0.33\textwidth]{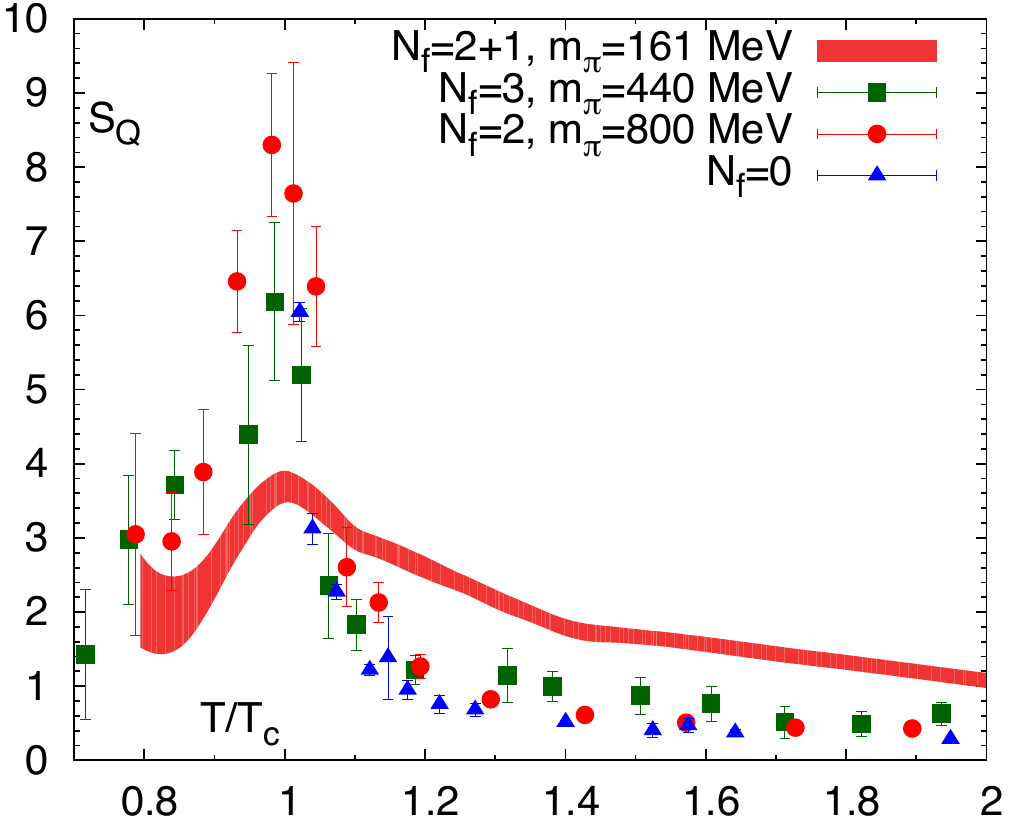}~~\includegraphics[width=0.33\textwidth]{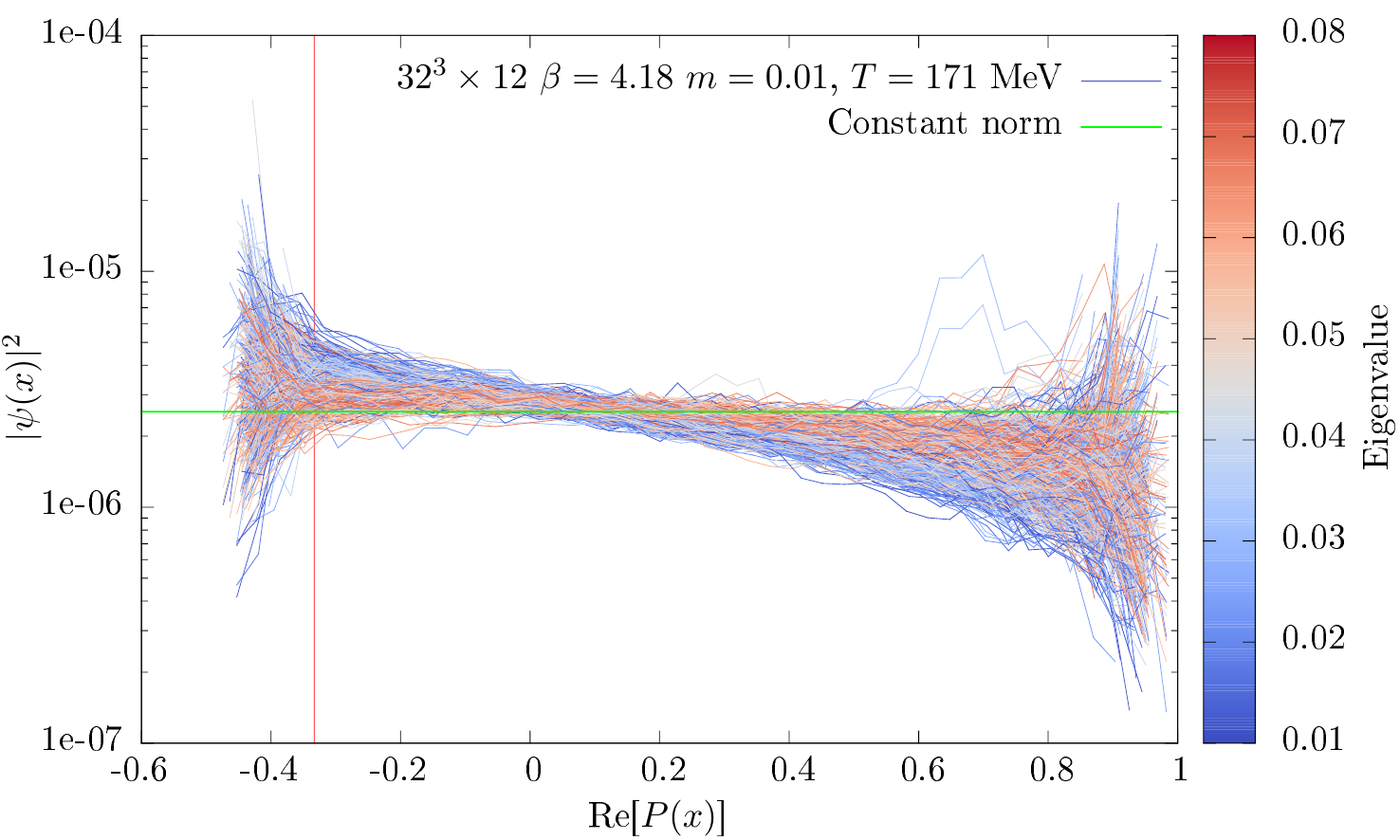}~~~~~~\includegraphics[width=0.33\textwidth]{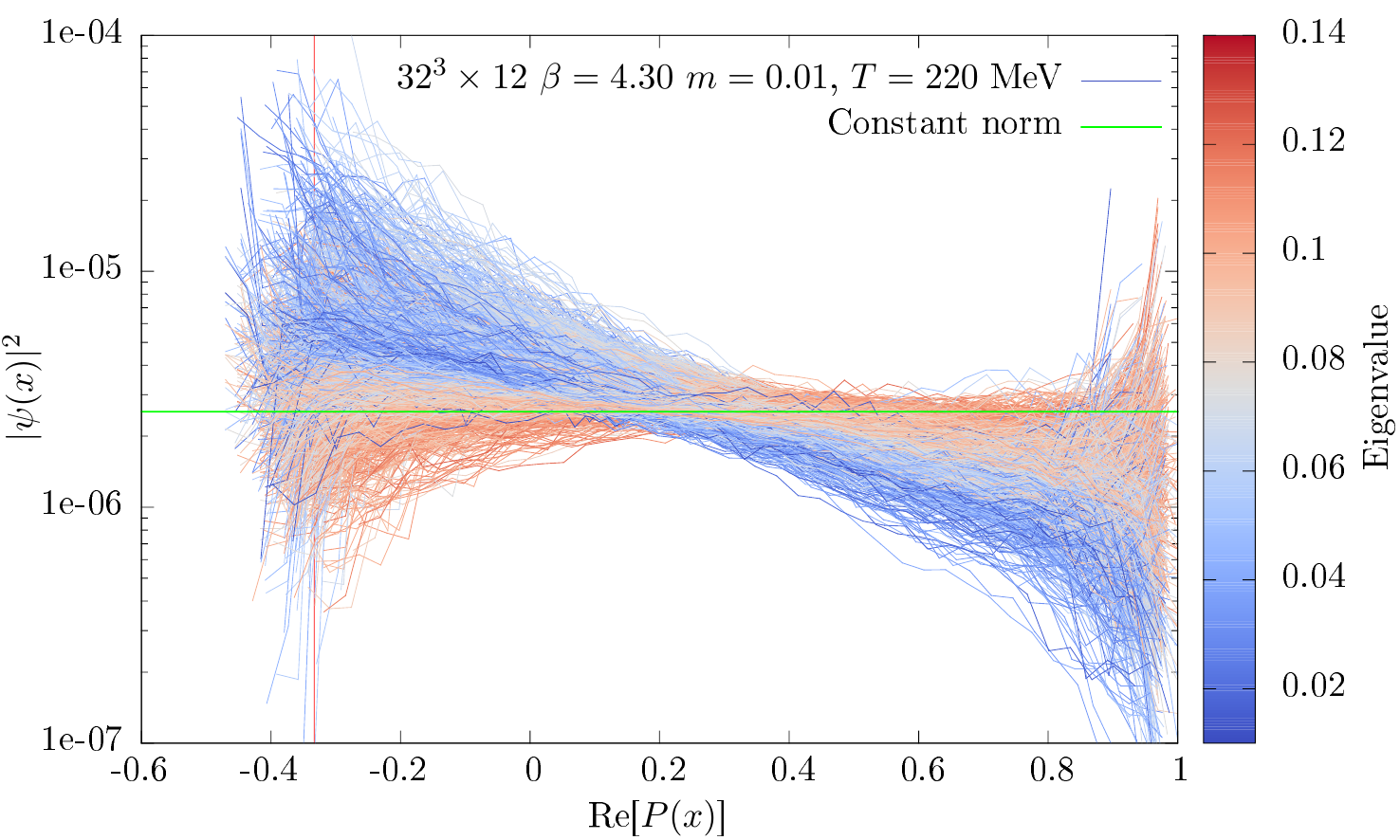}~\\
\end{center}
\caption{Left: Temperature dependence of a single-quark entropy~\cite{Bazavov:2016uvm}. The $x$ axis is rescaled by the corresponding lattice result for the transition temperature, i.e. here $T_c=153$, 193, 200 and 270 MeV for $N_f=2+1$, $N_f=3$, $N_f=2$ and $N_f=0$ (quenched) QCD, respectively. The red band shows the continuum extrapolated results based on simulations using HISQ fermions on $N_\tau=6$, 8, 10 and 12 lattices.  Middle and Right: The local square norm of the Dirac Eigenmode $|\psi(x)|^2$ in correlation with the real part of the trace of the local Polyakov loop $Re[P(x)]$ obtained from simulations using Domain Wall fermions~\cite{Cossu:2016scb}. Every single line corresponds to one eigenmode on one configuration. The color of the line represents the location of the eigenmode in the Dirac eigenvalue spectrum as shown in the contour bar. The green horizontal line represents a normalized flat mode where there is no correlation between the eigenmode and  the Polyakov loop. The vertical red line labels the location of $Re[P(x)]=1/3$.}
\label{fig:transitions}
\end{figure}

At nonzero quark masses there is no true phase transition in $N_f=2+1$ (2) QCD which may also due to the fact that binding energy, size and mass etc vary from particles.
It has been shown in Ref.~\cite{Bazavov:2013dta,Bazavov:2014yba} that heavier hadrons, e.g. open charm and open strange hadrons get deconfined in the chiral crossover temperature region. 
While it is found that in the SU(3) gauge theory susceptibilities of the imaginary and the real parts of the Polyakov loop can be used to probe the deconfinement phase transition~\cite{Lo:2013etb},
recently in the case of dynamic QCD the single-quark entropy $S_Q$ was introduced based on the Polyakov loop $P(x)$ as follows
\be
S_Q = -\frac{\partial F_Q}{\partial T},~~~~~F_Q = - T \log\Big(\langle P(x)\rangle\Big).
\ee
The temperature dependence of $S_Q$ is shown in the left panel of Fig.~\ref{fig:transitions}. By rescaling the temperature with the corresponding transition temperature $T_c$  $S_Q$
peaks at temperature just around $T_c$. This suggests that there could be possible connections between the deconfinment and chiral aspects of the QCD transition.
Although the connection between the Polyakov loop susceptibilities and the Dirac eigenvalue spectrum is found to be weak~\cite{Doi:2015kje,Suganuma:2016lnt}, the correlation between the local Polyakov loop
and the lowest Dirac eigenmode is clearly obsersved in the middle and right panels of Fig.~\ref{fig:transitions}~\cite{Cossu:2016scb}. At temperature below $T_c$, i.e. $T$=171 MeV, both high and low eigenmodes are delocalized and the local square norm of the Dirac eigenmode $|\psi(x)|^2$ has mild dependence on the real part of the trace of the local Polyakov loop $ReP(x)$. This is in contrast with the case at the temperature larger than $T_c$, i.e. $T$=220 MeV.
There the high modes remain almost unaffected by the temperature and the low modes surely become localized in the area about $ReP(x)=1/3$.

%

\section{QCD thermodynamics at high temperature}
\label{sec:Thermodynamics}

\subsection{Equation of State of $N_f=2+1$ and $N_f=0$~QCD}
\label{sec:EoS}

\begin{figure}[htp!]
\begin{center}
~\includegraphics[width=0.4\textwidth]{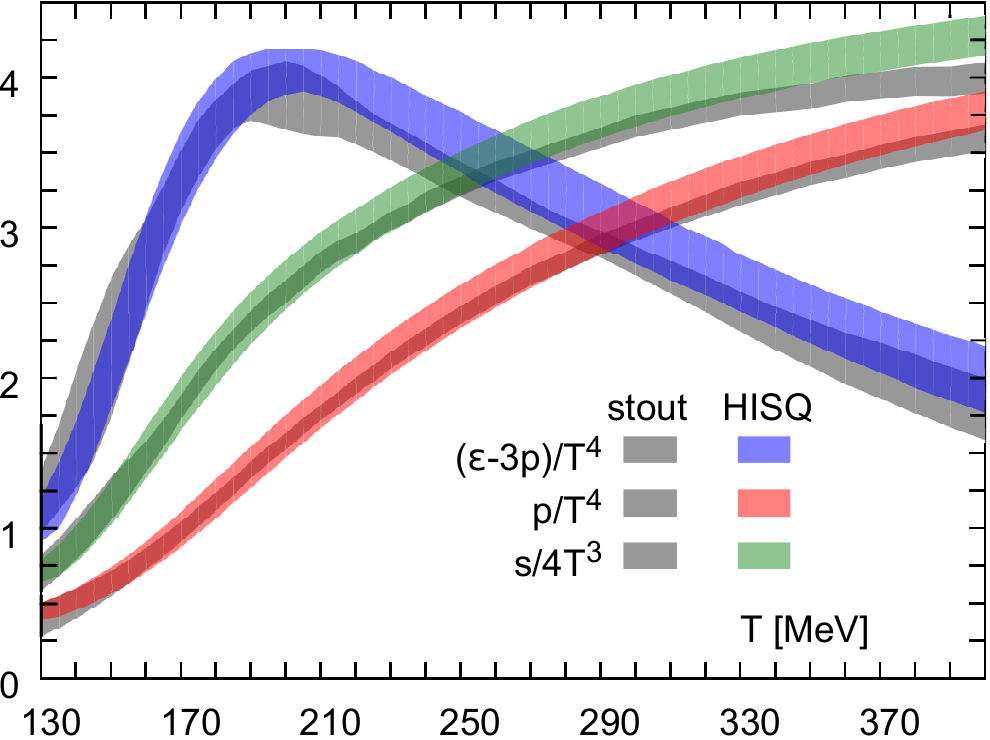}~~\includegraphics[width=0.4\textwidth]{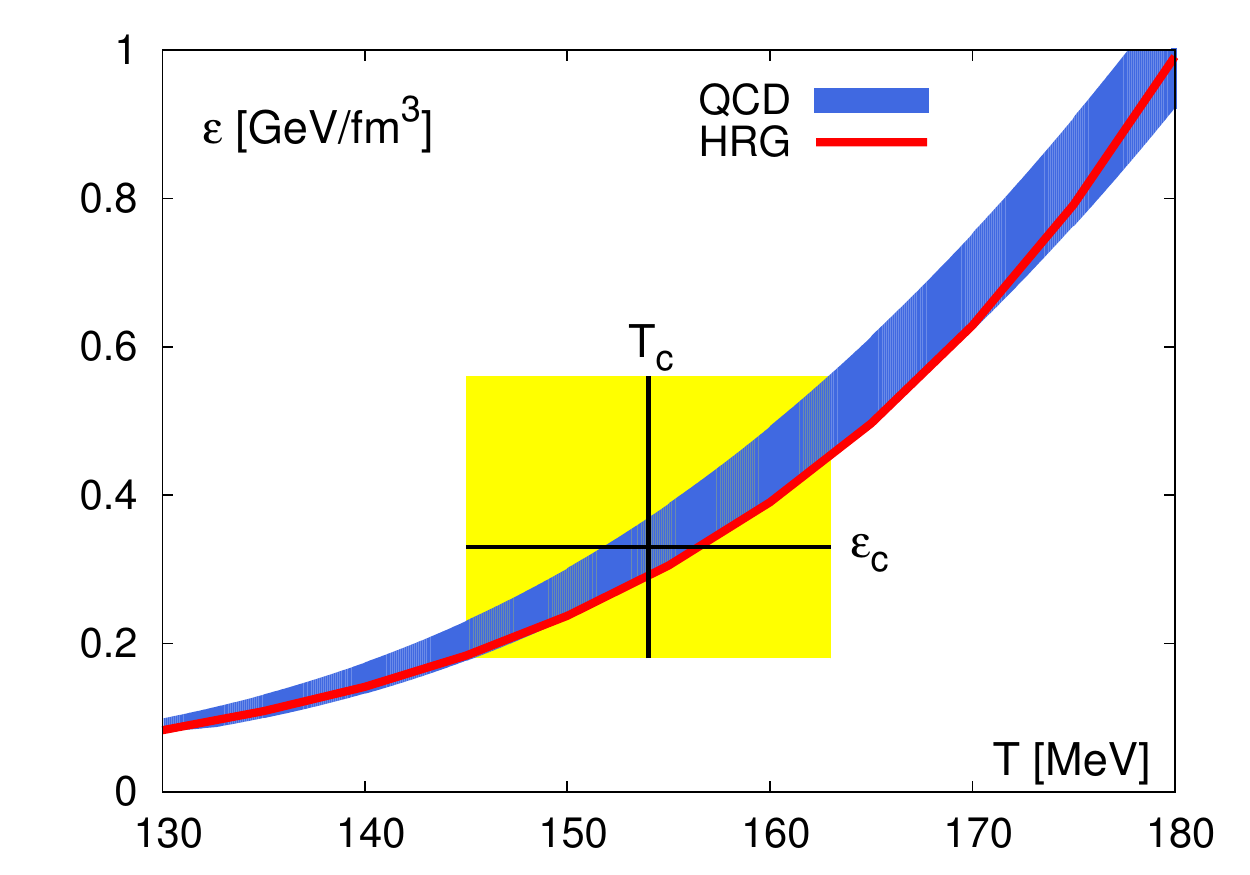}
\end{center}
\caption{Left: Continuum extrapolated results for equation of state of $N_f=2+1$ QCD obtained using HISQ~\cite{Bazavov:2014pvz} and stout fermions~\cite{Borsanyi:2013bia}. Right: The critical energy density $\epsilon_c$ in $N_f$=2+1 QCD. The band gives the continuum extrapolated result for the energy density taken from Ref.~\cite{Bazavov:2014pvz}.}
\label{fig:EoS}
\end{figure}

One of the milestones in the lattice QCD computation at finite temperature is the conclusive continuum extrapolated results for the equation of state (EoS) of $N_f=2+1$ QCD
with physical strange quark and light quark masses. The continuum extrapolated results on trace anomaly, pressure and entropy density from simulations using HISQ~\cite{Bazavov:2014pvz} and stout fermions~\cite{Borsanyi:2013bia}, as shown in the left panel of Fig.~\ref{fig:EoS}, agree with each other. In the right panel the critical energy density $\epsilon_c$ where chiral crossover transition happens is shown, i.e. 0.18 $<\epsilon_c < $ 0.5 GeV/fm$^3$. Note that $\epsilon_c$ is at the same order with the energy density of the nuclear matter, $\epsilon^{nuclear\,matter}\simeq 0.15$ GeV/fm$^3$ and the energy density inside a nucleon $\epsilon^{nuclear\,matter}\simeq 0.45$ GeV/fm$^3$ assuming the radius of nucleon to be  $\simeq$0.8 fm.
These mentioned calculations are based on the so-called T-integration method~\cite{Boyd:1996bx} and they are rather time-consuming as a subtraction of the 
zero temperature contributions is needed to eliminate divergent vacuum contributions. An alternative approach which is free of this issue is exploited in Refs.~\cite{Giusti:2012yj,Giusti:2014ila,Giusti:2016iqr}.
The general idea is to extract the entropy from the off-diagonal elements of the energy-momentum tensor which can be computed from ensembles defining in a moving frame.  The other thermodynamic quantities can be computed from thermodynamic relations. The application of this method is now only to the SU(3) gauge theory and the very recent results on EoS are presented in this conference~\cite{Giusti:2016wsf}. The obtained trace anomaly and pressure are quite consistent with results from the conventional T-integration method.
By going to very high temperature, i.e. up to  1000 $T_c$, the simulations become challenging as the physical spatial extent $L=a N_\sigma=N_\sigma/(TN_\tau)$ should not shrink as $T$ increases. This requires a larger ratio $N_\sigma/N_\tau$ at higher temperatures.

Another interesting method to compute EoS is to use the gradient flow~\cite{Luscher:2010iy,Luscher:2011bx,Luscher:2013vga}. The method to calculate the energy-momentum tensor in the SU(3) gauge theory has been proposed in Ref.~\cite{Suzuki:2013gza}. It is observed in the lattice QCD computations that the signal-to-noise ratio of bulk properties, e.g. pressure and energy density obtained using the gradient flow 
is significantly reduced. And these bulk quantities computed by the FlowQCD collaboration are consistent with the results obtained from high precision T-integration results~\cite{Asakawa:2013laa,Kitazawa:2016dsl}. The method has been extended to full QCD~\cite{Makino:2014taa} and the first numerical results for $N_f=$2+1 QCD with $O(a)$-improved Wilson quarks and pion mass larger than the physical one are reported in this conference~\cite{Taniguchi:2016ofw,Kanaya:2016rkt}.  
This method is promising also for the further calculations of energy-momentum tensor related quantities, e.g. viscosities etc~\cite{Suzuki:2016ytc}.

\subsection{Equation of State including contributions from charm quarks}
\label{sec:EoScharm}

Introducing heavy quarks, i.e. charm and bottom quarks on the lattice requires finer lattice spacing so that $a m_{\rm HQ}$<1.
Earlier computations on equation of state including charm quarks have been carried out by the MILC~\cite{Bazavov:2012kf,Bazavov:2013pra}, Wuppertal-Budapest~\cite{Ratti:2013uta} and ETM collaborations~\cite{Burger:2015xda}. In Ref.~\cite{Bazavov:2013pra} the trace anomaly was obtained using HISQ fermions with $m_\pi\approx$300 MeV on $N_\tau=6$, 8 and 10 lattices. It was found that the charm contribution to the trace anomaly becomes non-negligible starting at $T\approx$300 MeV and reaches its maxima at $T\approx$500 MeV.  The non-negligible contribution from charm quarks was also found at the temperature above the chiral crossover temperature employing twisted Wilson fermions and the fixed-scale approach with $m_\pi\approx370$ MeV on $N_\tau=3,4,\cdots24$ lattices~\cite{Burger:2015xda}.

\begin{figure}
\begin{center}
~\includegraphics[width=0.45\textwidth]{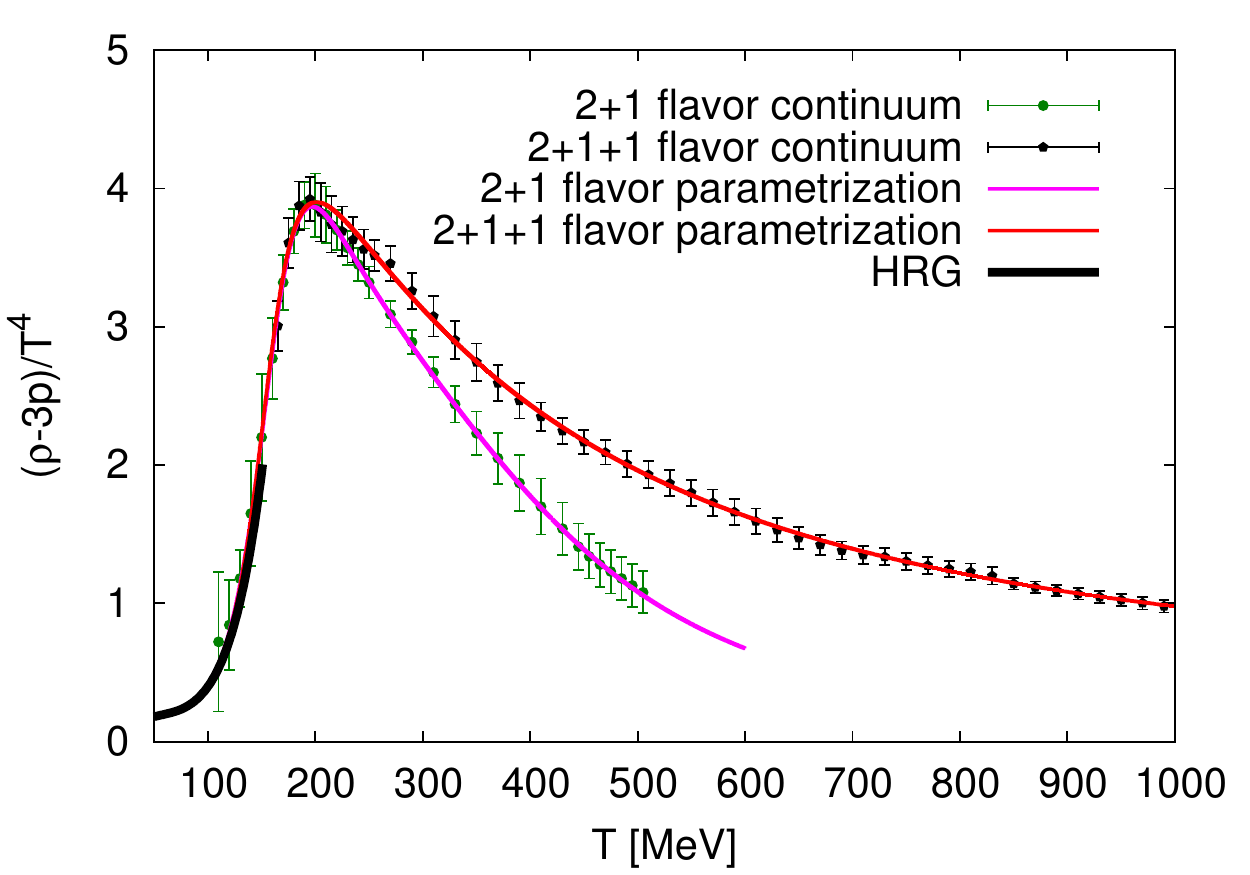}~~\includegraphics[width=0.45\textwidth]{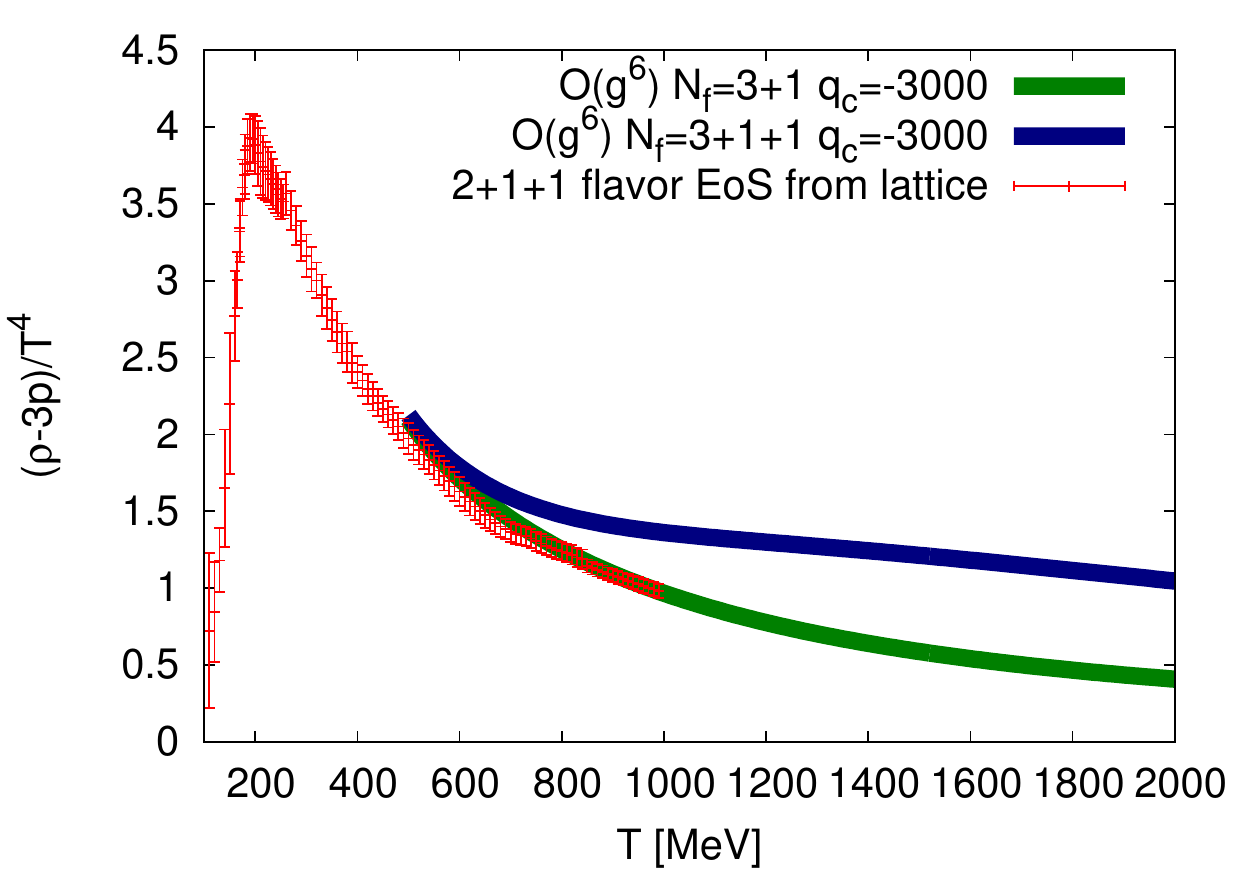}
\end{center}
\caption{Left: Continuum extrapolated results for trace anomaly of $N_f=2+1$ and $N_f=2+1+1$ QCD based on simulations using 4stout fermions with physical charm, strange and light quark masses on $N_\tau=6,$ 8, 10 and 12 lattices. Right: Reconstructed trace anomaly of $N_f=3+1+1$ QCD based on the lattice simulations of $N_f=2+1+1$ QCD and the perturbative QCD calculations up to $\mathcal{O}(g^6)$. Figures are taken from Ref.~\cite{Borsanyi:2016ksw}.}
\label{fig:EoScharm}
\end{figure}

Very recently the Wuppertal-Budapest collaboration updated the EoS results of $N_f=2+1+1$ QCD using 4stout fermions with physical charm, strange and light quark masses on $N_\tau=6,$ 8, 10 and 12 lattices~\cite{Borsanyi:2016ksw}.
Continuum extrapolated results for trace anomaly of $N_f=2+1+1$ QCD up to $T=1$ GeV are shown in the left panel of Fig.~\ref{fig:EoScharm}.  Large aspect ratios, i.e. $N_s/N_\tau\sim6$  are used to control the volume effects in high temperature region. By comparing to the trace anomaly of $N_f=2+1$ QCD it is clearly seen that the effect of the charm quark starts to be relevant at $T\approx$ 300 MeV. The effect of bottom quark is expected to appear at even higher temperature and the perturbation theory could then be applicable. Based on the analytic formula of pQCD up to $\mathcal{O}(g^6\ln g)$~\cite{Kajantie:2002wa} and  a one-parameter ($q_c$) fit of the numerical including the $\mathcal{O}(g^6)$ term~\cite{Hietanen:2008tv} with $m_c(m_c)=1.29$ GeV, the trace anomaly was extrapolated from lattice data for $N_f=2+1+1$ QCD to those for $N_f=3+1+1$ QCD with $m_b(m_b)$=4.18 GeV in a broader temperature window as shown in the right plane of Fig.~\ref{fig:EoScharm}. The contributions to pressure and energy density from photons, neutrinos, charged leptons as well as electroweak sectors are also added up in Ref.~\cite{Borsanyi:2016ksw} and the resulting EoS could then be relevant for the usage in cosmology.

\subsection{Topological susceptibility at finite temperature}
\label{sec:topsus}

The topological susceptibility of QCD $\chi_t$ is one of the most important observables that can be used to study the QCD vacuum properties at finite temperature, e.g. whether dyons exists around the chiral
crossover temperature and when a dilute gas of instantons can be a good description of the QCD vacuum etc~\cite{Gross:1980br}. Recently understanding $\chi_t$ becomes intensively investigated on the lattice~\cite{Berkowitz:2015aua,
Kitano:2015fla,Borsanyi:2015cka,Petreczky:2016vrs,Bonati:2015vqz,Taniguchi:2016tjc,Borsanyi:2016ksw} also due to its cosmological importance. 
The topological susceptibility is proportional to the squared mass of the axion which might be a possible candidate of dark matter~\cite{Marsh:2015xka}. The computation of the topological susceptibility on the lattice, however, is highly non-trivial~\cite{Vicari:2008jw}. 
Firstly the index theorem is not uniquely defined on the lattice and careful continuum extrapolations are needed. Secondly from the dilute instanton gas model~\cite{Gross:1980br} 
the topological susceptibility decreases with temperature with a power law $\chi_t~\sim T^{-\gamma}$ and $\gamma$ is found to be $6+2N_f/3$ by including quantum fluctuations to
the leading order in strong coupling constant over classical instanton action~\cite{Ringwald:1999ze}. This power-law decreasing behavior of $\chi_t$ with $T$
indicates that the weight of the configurations with nonzero topology decreases with the same power in $T$. Thus statistics of Monte Carlo simulations should be increased accordingly. 
Moreover the topological freezing which prevents the tunneling between topological sectors makes the computation more difficult.

\begin{figure}
\begin{center}
~\includegraphics[width=0.5\textwidth]{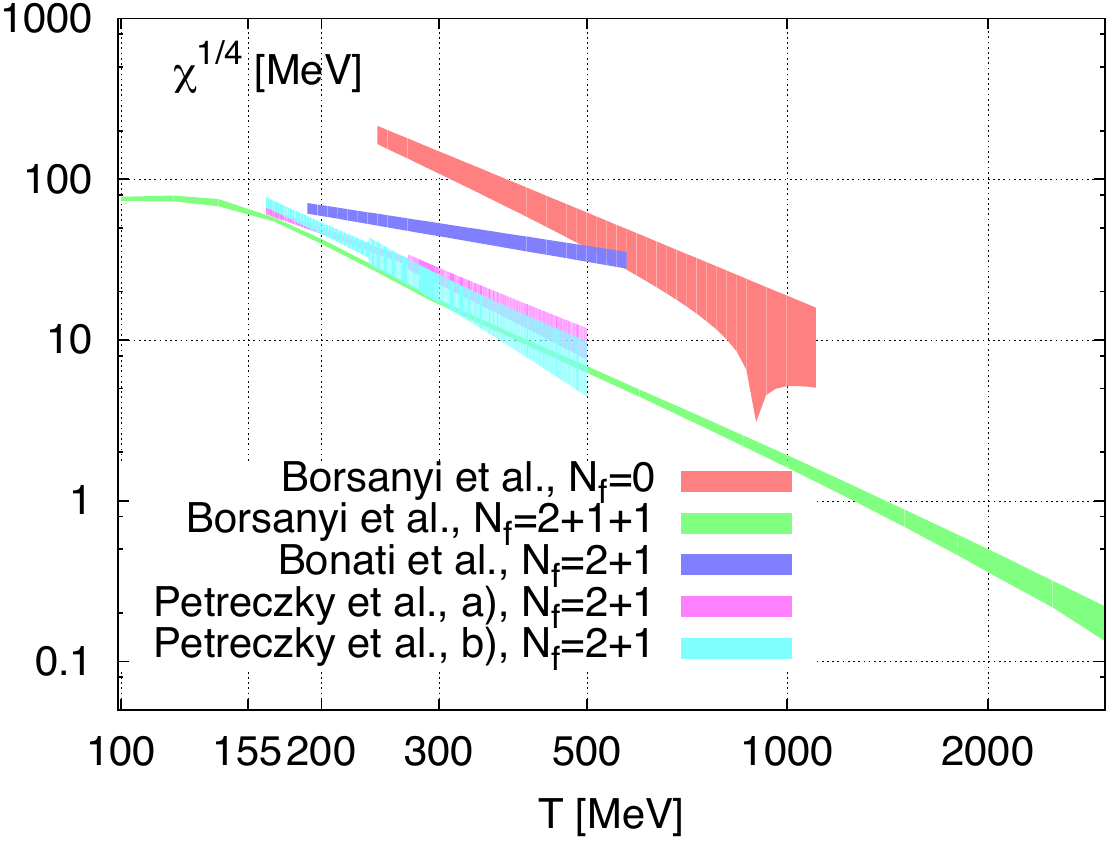}
\end{center}
\caption{Continuum extrapolated results for topological susceptibility in quenched QCD~\cite{Borsanyi:2015cka}, $N_f=2+1$ QCD with the gluonic definition of the topological charge~\cite{Bonati:2015vqz}, $N_f=2+1$ QCD with a) gluonic definition and b) $\chi_t=m^2\chi_{disc}$~\cite{Petreczky:2016vrs}, and $N_f=2+1+1$ QCD with fixed topology sector method~\cite{Borsanyi:2016ksw}.}
\label{fig:topsus}
\end{figure}

To circumvent the above mentioned issues several strategies are proposed, including the usage of the topological charge density~\cite{ Bautista:2015yza}, assuming a Gaussian distributions of 
topological charges in sub-volumes~\cite{Bietenholz:2015rsa}, artificial decreasing the weight of the trivial sector~\cite{Kitano:2015fla}, gradient flow~\cite{Taniguchi:2016tjc} and the fixed topology sector~\cite{Frison:2016vuc,Borsanyi:2016ksw}. Recently $\chi_t$ has been computed  on the lattice in quenched QCD~\cite{Berkowitz:2015aua,Kitano:2015fla,Borsanyi:2015cka},
$N_f=2+1$ QCD having physical pion masses with stout~\cite{Bonati:2015vqz}, HISQ~\cite{Petreczky:2016vrs},  Wilson fermions~\cite{Taniguchi:2016tjc} and HISQ/overlap fermions~\cite{Sharma:2016cmz}
as well as $N_f=2+1+1$ QCD~\cite{Borsanyi:2016ksw} using stout/overlap fermions. Continuum extrapolated results for $\chi_t$ in quenched and full QCD by three groups~\cite{Borsanyi:2015cka,Bonati:2015vqz,Petreczky:2016vrs,Borsanyi:2016ksw} are shown in Fig.~\ref{fig:topsus}. Further detailed studies are needed to resolve the discrepancies among various groups.

\section{QCD Equation of State at nonzero baryon number density}
\label{sec:EoSMu}

QCD equation of state (EoS) at nonzero baron density is indispensable for the hydrodynamic modeling of the conditions met in the current/future heavy ion 
experiment program at low beam energies. In the current beam energy scan program at RHIC the  freeze-out baryon chemical potential $\mu_B$ is within the range of 20 to 400 MeV and the freeze-out temperature 140 $\lesssim T\lesssim$170 MeV~\cite{Cleymans:2005xv,Andronic:2008gu}. This thus requires an EoS at nonzero baryon density for 0$\lesssim \mu_B/T\lesssim$3.
At nonzero baryon chemical potential lattice QCD computations are hindered by the so-called sign problem.
To circumvent this problem the Taylor expansion~\cite{Allton:2002zi,Gavai:2003mf} and Imaginary $\mu$~\cite{deForcrand:2002hgr,DElia:2002tig} methods are commonly used to compute QCD thermodynamics at the physical pion mass.  
In the framework of the Taylor expansion method, the difference of the pressure at nonzero and vanishing values of $\mu_B$
can be expressed as the following Taylor series
\begin{eqnarray}
\frac{P(T,\mu_B)-P(T,0)}{T^4} &=&  
\sum_{n=1}^\infty \frac{\chi_{2n}^{B}(T)}{(2n)!}
\left(\frac{\mu_B}{T}\right)^{2n}  \\
&=& \frac{1}{2} \chi_{2}^B(T) \hmu_B^2 \left( 1+ \frac{1}{12}
\frac{\chi_{4}^B(T)}{\chi_{2}^B(T)} \hmu_B^2
+ \frac{1}{360}\frac{\chi_{6}^B(T)}{\chi_{2}^B(T)} \hmu_B^4+\; ... \right)
\label{eq:PmuB}.
\end{eqnarray} 
where $\chi_{2n}^{B}(T)=\frac{\partial P(T,\hmu)/T^4}{\partial \hmu_B^{2n}}|_{\hmu=0}$ is the $2n$th order Taylor expansion coefficient and can be computed directly at $\hat{\mu}_B=\mu_B/T=0$ on the lattice. Here the odd order of the coefficient vanishes due
to the C-symmetry of the QCD action. Once the expansion coefficients are obtained  all other bulk quantities can be obtained according to thermodynamic relations.
To provide an EoS that is reliable up to $\mu_B/T\approx$3
at least 8th order of the Taylor expansion coefficients are needed.  And the main challenge in lattice QCD computations of the Taylor expansion coefficients is that the number of the fermion matrix inversions increases significantly at such a high order derivatives and so does the noise-to-signal ratio.  While in the simulations at imaginary chemical potentials, there is no sign problem and thermodynamic quantities can be computed directly. However, an analytic continuation to real chemical potentials from imaginary chemical potentials is needed and is one of the main sources of uncertainties.

\begin{figure}
\begin{center}
~\includegraphics[width=0.45\textwidth]{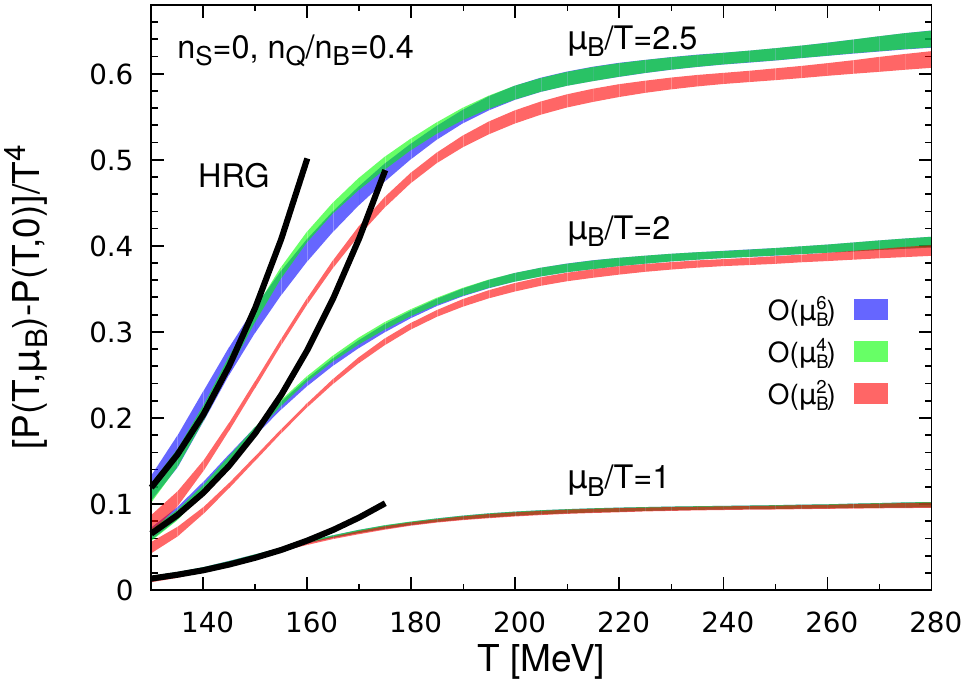}~~\includegraphics[width=0.45\textwidth]{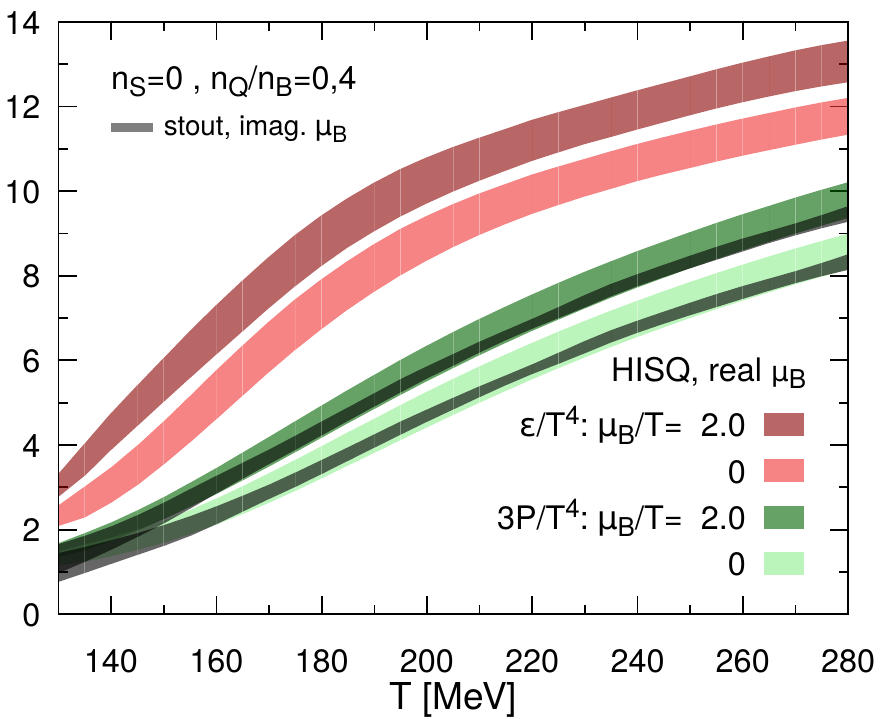}
\end{center}
\caption{Left: Pressure differences of $N_f=2+1$ QCD for $\mu_B/T$=1, 2 and 2.5 truncated at the orders of $\hmu_B^2$, $\hmu_B^4$ and $\hmu_B^6$ obtained from the Taylor expansion method using HISQ fermions~\cite{Bazavov:2017dus}.  The EoS as shown is reliable at $\mu_B/T\lesssim 2$, i.e. beam energy $\sqrt{s_{\rm NN}}\gtrsim 12$ GeV~\cite{Bazavov:2017dus}.
Right: Consent of EoS in the continuum limit at both zero and nonzero $\mu_B$ obtained from analytic continuation from imaginary chemical potentials using stout fermions~\cite{Gunther:2016vcp} and the Taylor expansion method using HISQ fermions~\cite{Bazavov:2017dus}. The total energy density (upper two curves) and three times pressure (lower two curves) of $N_f=2+1$ QCD at $\mu_B=0$ and 2 obtained from simulations using HISQ fermions and the Taylor expansion method are shown~\cite{Bazavov:2017dus}. The dark lines show the results obtained from analytic continuation with sixth order polynomials in $\hat{\mu}_B$ from simulations at imaginary chemical potentials using stout fermions~\cite{Gunther:2016vcp}. Figures are taken from Ref.~\cite{Bazavov:2017dus}.}
\label{fig:EoSMu}
\end{figure}

E. Laermann from the Bielefeld-BNL-CCNU collaboration presented in this conference the Equation of State of $N_f=2+1$ QCD evaluated
up to six order of baryon chemical potential $\hat{\mu}_B$ using the Taylor Expansion Method~\cite{Bazavov:2017dus}. The simulations are carried out using HISQ fermions on $N_\tau=$6, 8, 12 and 16 lattices. To have accurate determinations of coefficients
a large number of gauge configurations, $\sim10^5$ for each $T$, and the linear $\mu$ method for high order coefficients~\cite{Gavai:2011uk,Gavai:2014lia} are used in the computation.
To meet the strangeness neutral case, i.e. mean value of the strangeness $n_S$ is zero and the ratio of the mean values of electrical charge and baryon number $n_Q/n_B$ is 0.4, in the heavy ion collision experiment,
the dependence of $\mu_B$ on the electrical charge $\mu_Q$ and strangeness chemical potential $\mu_S$  has to be taken into account. This is done by re-expressing the Taylor expansion coefficients also in terms of pressure derivates of $\mu_S$ and $\mu_Q$ computed at the vanishing values of potentials. As shown in the left panel of ~Fig.~\ref{fig:EoSMu} the computation of the pressure difference meeting the strangeness neutral case between nonzero and zero $\mu_B$ is reliable at $\mu_B/T\lesssim$ 2 with the expansion truncated at the 6th order of Taylor expansion coefficients.

 J. G\"unther from the Wuppertal-Budapest collaboration also presented their recent study on equation of state of $N_f=2+1$ QCD at nonzero $\mu_B$~\cite{Gunther:2016vcp}. The EoS is obtained from analytic continuation from simulations at imaginary chemical potentials using stout fermions on $N_\tau=10,$ 12 and 16 lattices. The conditions met in the heavy ion collisions, i.e. $n_S$=0 and $n_Q/n_B$=0.4, are imposed directly in the lattice simulation~\cite{ Bellwied:2015rza}. The uncertainties from analytic continuation are studied with various forms. The obtained energy density and three times pressure are shown as the dark lines in the right panel of Fig.~\ref{fig:EoSMu}. It is clearly seen that the results from two groups are in good agreement.

\section{QCD phase structure at nonzero chemical potentials}
\label{sec:PhaseMu}

\begin{figure}
\begin{center}
~\includegraphics[width=0.45\textwidth]{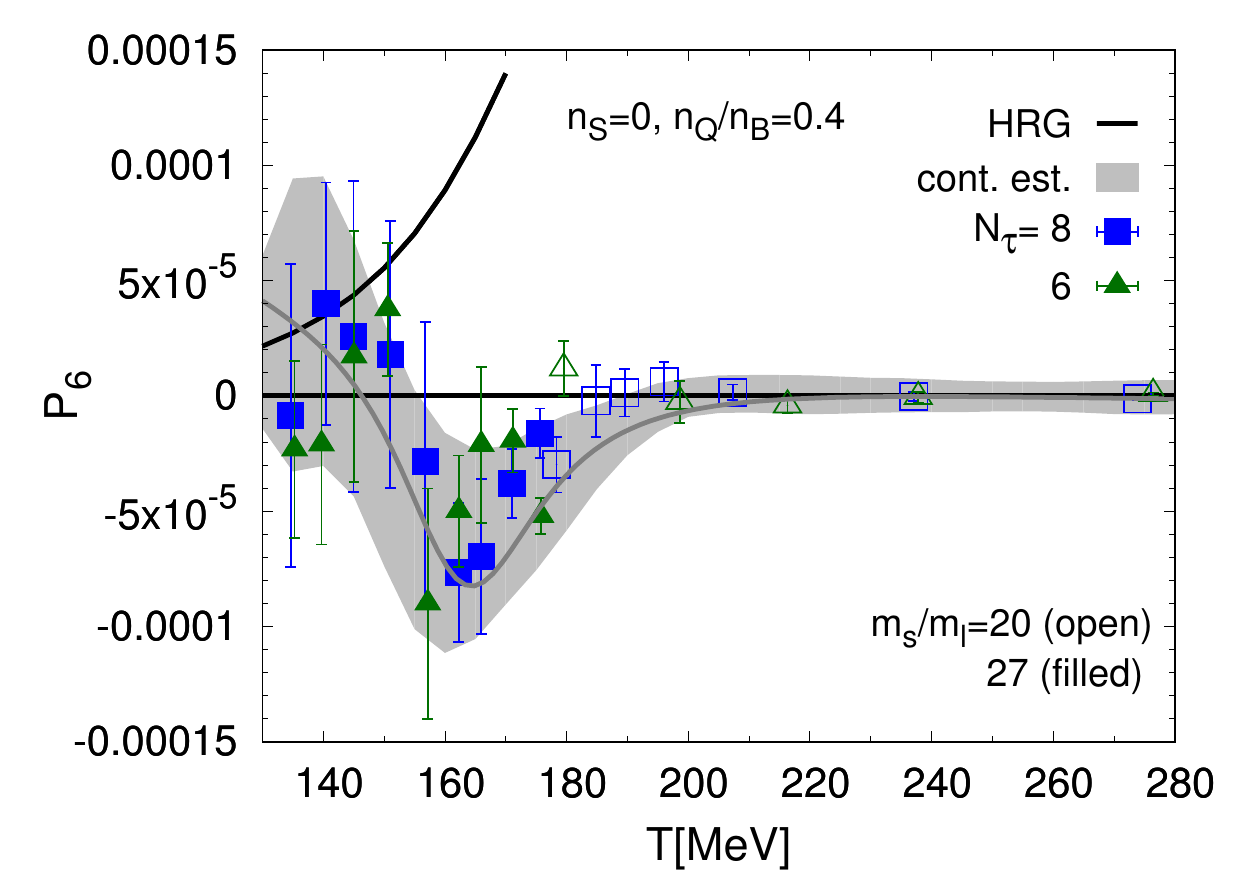}~~\includegraphics[width=0.45\textwidth]{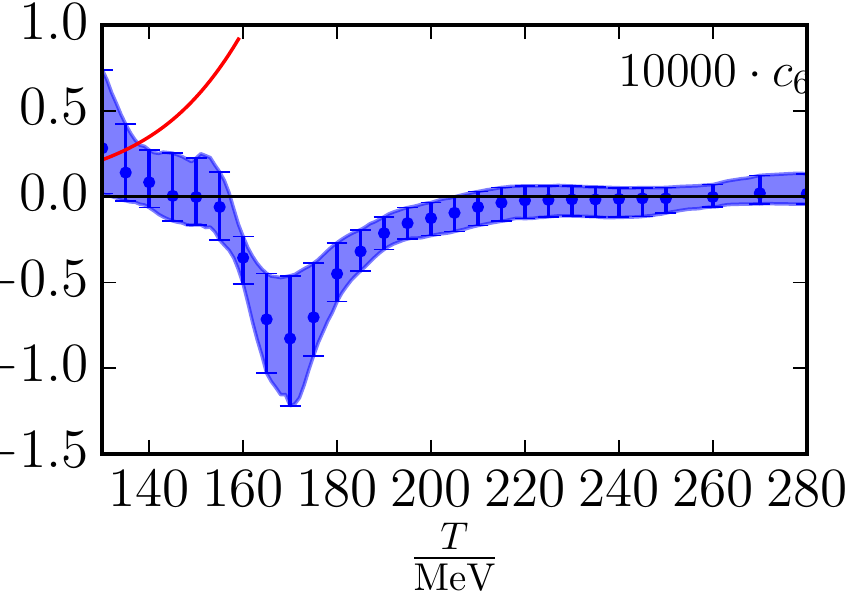}
\end{center}
\caption{Six order Taylor expansion coefficient of the QCD pressure ($P_6$ and $c_6$) obtained from simulations using HISQ fermions~\cite{Bazavov:2017dus} and stout fermions at imaginary chemical potentials~\cite{Gunther:2016vcp}.}
\label{fig:6thExpCoeff}
\end{figure}

Searching the critical point in the QCD $T-\mu_B$ diagram is one of the main goals in the Beam Energy Scan program at RHIC~\cite{Luo:2017faz}.
One of the important probes is the fluctuations of conserved charges which is computable and measurable on the lattice and in the heavy ion experiment, respectively.
The full understanding of the non-monotonic behavior of the ratio, i.e.  4th order to 2nd order proton number fluctuations, $\kappa_p\sigma^2_p=\chi_{4,\mu}^B/\chi_{2,\mu}^B$, observed in the STAR experiment certainly requires computations of higher-than-6th order Taylor expansion coefficients.
As mentioned in the previous section the computation of high order Taylor expansion coefficients, from which the fluctuations of conserved charges at nonzero $\mu_B$ can 
be constructed, is a great challenge. While the continuum extrapolated results for second and fourth order Taylor expansion coefficients have been computed~\cite{Borsanyi:2011sw,Bazavov:2012jq,Bazavov:2013uja,Ding:2015fca,Bellwied:2015lba}, the 6th order Taylor expansion coefficient of pressure in the continuum limit
are published very recently~\cite{Bazavov:2017dus,Gunther:2016vcp} as shown in Fig.~\ref{fig:6thExpCoeff}. The striking behavior seen in the $T$ dependence of the coefficient is the 
flip of the sign at the temperature slightly larger than the chiral cross over temperature. This is comparable with the results obtained from 
Polyakov loop extended quark meson model having a global O(4) symmetry~\cite{Friman:2011pf}.
Together with the positivity of the 2nd and 4th order coefficients the decreasing behavior with beam energy of $\kappa_p\sigma^2_p$ measured in BES can be qualitatively described assuming the same condition on lattice can be imposed 
to the experiment data.

\begin{figure}
\begin{center}
\includegraphics[width=0.54\textwidth]{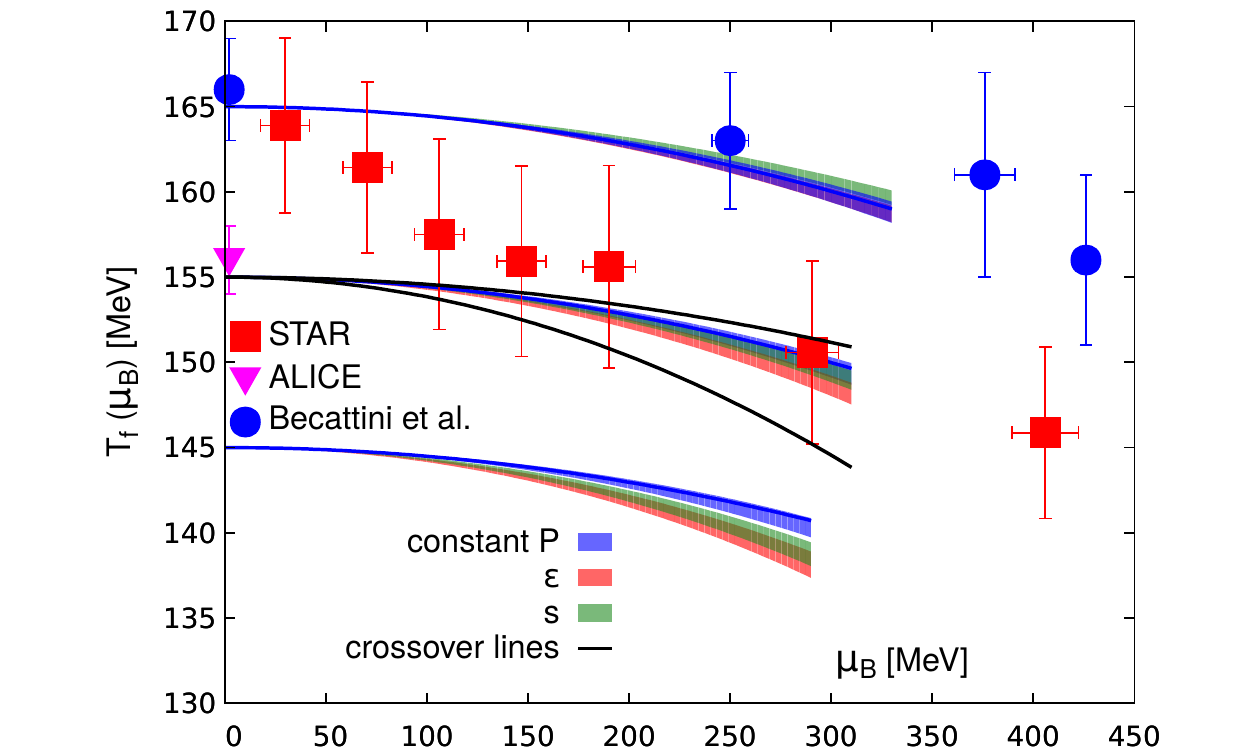}~\includegraphics[width=0.45\textwidth]{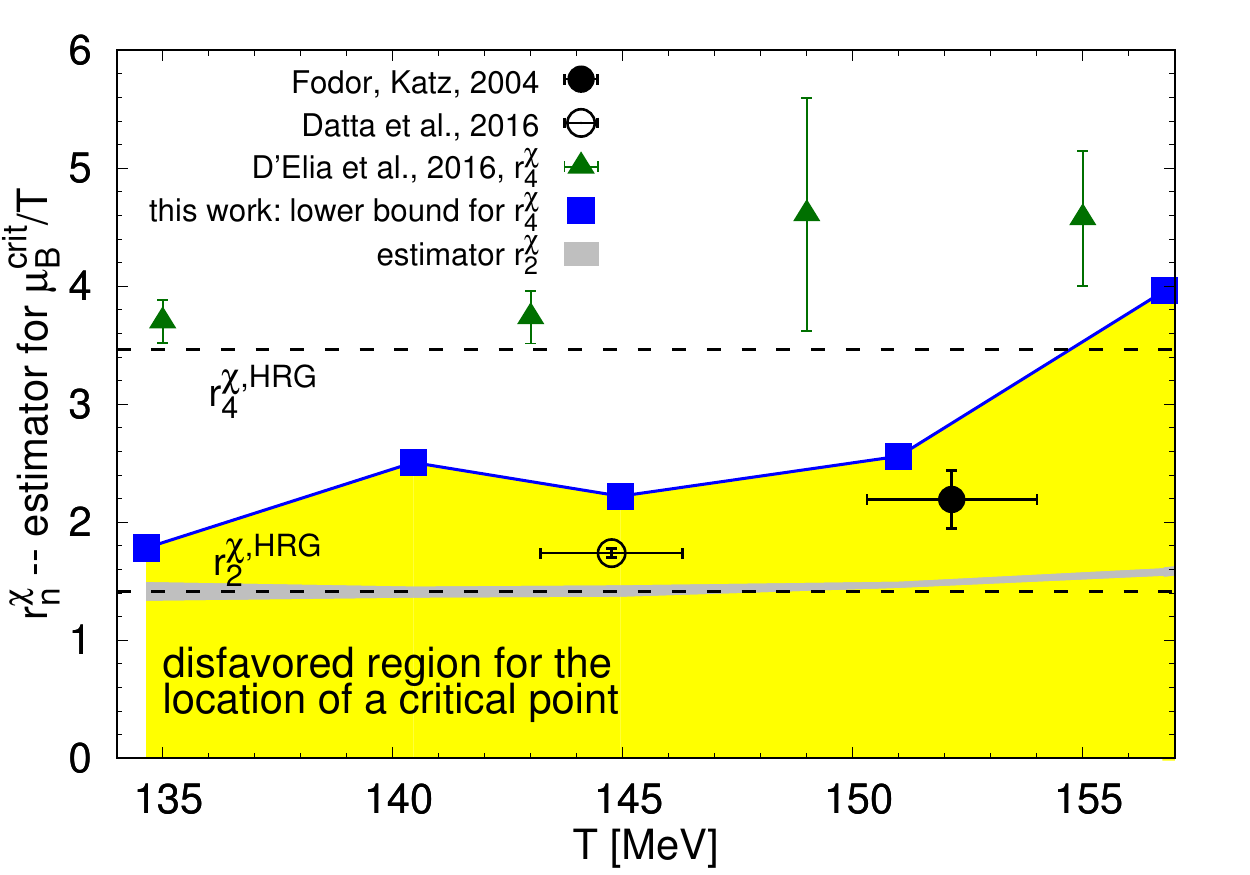}
\end{center}
\caption{ Left: Lines of constant pressure $P$, energy density $\epsilon$ and entropy density $s$ versus temperature in (2+1)-flavor QCD for three different initial sets of values fixed at $\mu_B$ = 0 and $T(\mu_B=0)$ = 145, 155 and 165 MeV, respectively. Data points show freeze-out temperatures determined by the STAR Collaboration in the BES at RHIC (squares)~\cite{Das:2014qca} and the ALICE Collaboration at the LHC (triangle)~\cite{Floris:2014pta}. The circles denote hadronization temperatures obtained by comparing experimental data on particle yields with a hadronization model calculation~\cite{Becattini:2016xct}. Also shown are two lines representing the current spread in determinations of the $\mu_B$-dependence of the QCD crossover transition line~\cite{Kaczmarek:2011zz,Endrodi:2011gv,Bonati:2014rfa,Cea:2015cya,Bellwied:2015rza}. Right:
The estimate of the radius of convergence $r_{2n}^{\chi}$ obtained from various groups.  Shown are lower
bounds for the estimator $r_4^{\chi}$ obtained in Ref.~\cite{Bazavov:2017dus} (squares)
and results for this estimator obtained from calculations with an
imaginary chemical potential (triangles)~\cite{DElia:2016jqh}.
Also shown are estimates for the location of the critical point
obtained from calculations with unimproved staggered fermions using
a reweighting technique~\cite{Fodor:2004nz} and Taylor expansions~\cite{Datta:2016ukp}.
In both cases results have been rescaled using $T_c=154$~MeV. Figures are taken from Ref.~\cite{Bazavov:2017dus}.}
\label{fig:PhaseMuB}
\end{figure}

It is also important to study the feasibility of detecting the critical point in the heavy ion experiment. As in the heavy ion experiment the freeze-out line is `measured` while the critical point is located on the crossover transition line.
While the curvature of the crossover line/chiral phase transition line can be naturally computed on the lattice~\cite{Kaczmarek:2011zz,Endrodi:2011gv,Bonati:2014rfa,Cea:2015cya,Bellwied:2015rza}, it was proposed that the freeze-out temperature and the curvature of the freeze-out line can also be extracted by comparing the experimental cumulant ratios with those computed on the lattice~\cite{Bazavov:2012vg,Bazavov:2014xya, Bazavov:2015zja}. The lattice computations on the parameterization of freeze out lines are also presented in Refs.~\cite{Borsanyi:2013hza,Bazavov:2014xya,Borsanyi:2014ewa,Bellwied:2016cpq}.  Additionally it has been argued quite successfully that lines on which certain thermodynamic observables or ratios thereof stay constant can describe the thermal conditions at the time of the chemical freeze-out in heavy ion collisions~\cite{Cleymans:1999st,Cleymans:2005xv}. Very recently the isentropic lines, i.e. with constant ratio of entropy to baryon number are presented~\cite{Gunther:2016vcp}. And the $\mu_B$ dependences of lines of constant pressure $P$, entropy $s$ and energy density $\epsilon$ are computed using the Taylor expansion method~\cite{Bazavov:2017dus}. The obtained curvatures are 
\begin{equation}
0.0064 \le \kappa_2^P  \le 0.0101 \;\; ,\;\; 
0.0087 \le \kappa_2^\epsilon  \le 0.012  \;\; ,\;\;
0.0074 \le \kappa_2^s  \le 0.011 \; .
\label{kappa2}
\end{equation}
These curvatures are comparable to the curvature of the crossover lines $0.0066\lesssim\kappa_{2}^{\rm crossover}\lesssim0.018$~\cite{Kaczmarek:2011zz,Endrodi:2011gv,Bonati:2014rfa,Cea:2015cya,Bellwied:2015rza}.
In the left panel of Fig.~\ref{fig:PhaseMuB} lines of constant pressure $P$, entropy $s$ and energy density $\epsilon$ as well as the crossover transition and freeze-out lines are shown. While an upper bound on the curvature of the freeze-out line $\kappa_2^{\rm freeze-out}\lesssim 0.011$ is obtained by comparing the lattice QCD results to the RHIC data~\cite{Bazavov:2015zja}, the freeze-out line obtained by the STAR collaboration, ALICE collaboration and a hadronization model are not consistent among each other~\cite{Das:2014qca,Floris:2014pta,Becattini:2016xct}.

The location of a possible critical point, on the other hand, has not been determined reliably yet from lattice QCD computations. The radius of convergence of the Taylor expansion series, $R$, can be used to estimate the possible location of
the critical point,
\be
R=\lim_{n\rightarrow\infty}r_{2n}^{\chi} = \left| \frac{2n (2n-1)\chi_{2n}^B}{\chi_{2n+2}^B} \right|^{1/2} \; ,
\ee
if all $\chi_{2n}^B$ stays asymptotically positive at real values of $\hat{\mu}_B$.
Here $\chi_{2n}^B$ is the $2n$th order derivatives of the pressure with respect to $\hat{\mu}_B$ evaluated at $\hat{\mu}_B=0$ as shown in Eq.~(\ref{eq:PmuB}).
The right panel of Fig.~\ref{fig:PhaseMuB} summarizes current values of $r_{2n}^{\chi}$ calculated from different groups.
From the plot it is concluded that a critical point at $\mu_B/T\lesssim$2 is strongly disfavored in the temperature range 135 MeV $\leq T\leq$ 155 MeV and its location at higher values of temperature seems to be ruled out~\cite{Bazavov:2017dus}.

Other than Taylor expansion and imaginary chemical potential methods many promising approaches, e.g. Lefschetz thimbles~\cite{Cristoforetti:2012su,Scorzato:2015qts}, complex Langevin simulations~\cite{Aarts:2009uq,Aarts:2017vrv}, density of sate methods~\cite{Langfeld:2012ah,Langfeld:2016kty}, canonical approaches~\cite{Nakamura:2015jra,Bornyakov:2016wld} have been proposed to solve the sign problem and then to study the QCD phase diagram at nonzero baryon number density directly. However, their applicability to full QCD have not been proven yet and more work needs to be done.

Recent studies on the QCD phase structure at finite isospin chemical potential $\mu_I$ have been presented by B. Brandt and G. Endrodi in this conference~\cite{Brandt:2016zdy}.
The sign problem is not present in QCD at nonzero $\mu_I$ so the conventional algorithms can be applied in the lattice QCD simulations. 
A novel method based on the singular value spectrum of the massive Dirac operator and a leading-order reweighting are shown to better determine the renormalized pion condensate.
Preliminary results for the phase boundary to the pion condensation phase and crossover line from chiral condensates are shown and the nature of transitions on these boundary lines are being pursued.

Updated studies on the Roberge-Weiss phase transition in $N_f=2+1$ QCD have also been presented by M. Mesiti~\cite{Bonati:2016pwz} in this conference.
The continuum extrapolated results on the Roberge-Weiss phase transition temperature $T_{RW}=$208 MeV is obtained by simulations using stout fermions with physical pion mass on $N_\tau$=4, 6, 8 and 10 lattices.
Based on the finite size scaling analyses the order of the transition is more likely to be second order belonging to 3d-Ising universality class rather than first order for $N_\tau=4$ and 6 lattices.
Studies on the Roberge-Weiss phase in $N_f=2$ QCD using standard Wilson fermions~\cite{Cuteri:2015qkq} and staggered fermions~\cite{Philipsen:2016swy} are also presented.

\vspace{-0.2cm}

\section*{Acknowledgements}

I would like to thank members of the Bielefeld-BNL-CCNU collaboration for the collaborations over the past years.
Thanks also go to people who sent me materials/plots before my talk: Pedro Bicudo, Jacques Bloch, Szabolcs Borsanyi, Bastian Brandt, Falk Bruckmann, Guido Cossu, Gergely Endrodi, Philippe de Forcrand, Jana G\"unther, Yoichi Iwasaki, Kazuyuki Kanaya, Sandor Katz, Masakiyo Kitazawa, Yu Maezawa, Atsushi Nakamura, Yoshifumi Nakamura, Alexander Rothkopf, Jonivar Skullerud, Yusuke Taniguchi, Helvio Vairinhos, Takashi Umeda although not all of their achievements are included in my proceedings/talk. 
%


\begin{thebibliography}{100}

\bibitem{Ding:2015ona}
H.-T. Ding, F.~Karsch and S.~Mukherjee, \emph{{Thermodynamics of
  strong-interaction matter from Lattice QCD}},
  \href{http://dx.doi.org/10.1142/S0218301315300076}{\emph{Int. J. Mod. Phys.}
  {\bf E24} (2015) 1530007}, [\href{https://arxiv.org/abs/1504.05274}{{\tt
  1504.05274}}].

\bibitem{Meyer:2011gj}
H.~B. Meyer, \emph{{Transport Properties of the Quark-Gluon Plasma: A Lattice
  QCD Perspective}},
  \href{http://dx.doi.org/10.1140/epja/i2011-11086-3}{\emph{Eur.Phys.J.} {\bf
  A47} (2011) 86}, [\href{https://arxiv.org/abs/1104.3708}{{\tt 1104.3708}}].

\bibitem{Schmidt:2017bjt}
C.~Schmidt and S.~Sharma, \emph{{The phase structure of QCD}},
  \href{https://arxiv.org/abs/1701.04707}{{\tt 1701.04707}}.

\bibitem{Sexty:2014dxa}
D.~Sexty, \emph{{New algorithms for finite density QCD}}, {\emph{PoS} {\bf
  Lattice2015} (2014) 016}, [\href{https://arxiv.org/abs/1410.8813}{{\tt
  1410.8813}}].

\bibitem{Meyer:2015wax}
H.~B. Meyer, \emph{{QCD at non-zero temperature from the lattice}}, {\emph{PoS}
  {\bf LATTICE2015} (2016) 014}, [\href{https://arxiv.org/abs/1512.06634}{{\tt
  1512.06634}}].

\bibitem{Borsanyi:2015axp}
S.~Bors{\'a}nyi, \emph{{Fluctuations at finite temperature and density}},
  {\emph{PoS} {\bf LATTICE2015} (2016) 015},
  [\href{https://arxiv.org/abs/1511.06541}{{\tt 1511.06541}}].

\bibitem{Bazavov:2015qsa}
A.~Bazavov, \emph{{Lattice QCD at Non-Zero Temperature}},  in
  \emph{{Proceedings, 32nd International Symposium on Lattice Field Theory
  (Lattice 2014): Brookhaven, NY, USA, June 23-28, 2014}}, 2015.
\newblock \href{https://arxiv.org/abs/1505.05543}{{\tt 1505.05543}}.

\bibitem{DElia:2015rwa}
M.~D'Elia, \emph{Lattice qcd with purely imaginary sources at zero and non-zero
  temperature}, {\emph{PoS} {\bf LATTICE2014} (2015) 020}.

\bibitem{Scorzato:2015qts}
L.~Scorzato, \emph{{The Lefschetz thimble and the sign problem}}, {\emph{PoS}
  {\bf LATTICE2015} (2016) 016}, [\href{https://arxiv.org/abs/1512.08039}{{\tt
  1512.08039}}].

\bibitem{Pisarski:1983ms}
R.~D. Pisarski and F.~Wilczek, \emph{{Remarks on the Chiral Phase Transition in
  Chromodynamics}},
  \href{http://dx.doi.org/10.1103/PhysRevD.29.338}{\emph{Phys.Rev.} {\bf D29}
  (1984) 338--341}.

\bibitem{Pelissetto:2013hqa}
A.~Pelissetto and E.~Vicari, \emph{{Relevance of the axial anomaly at the
  finite-temperature chiral transition in QCD}},
  \href{http://dx.doi.org/10.1103/PhysRevD.88.105018}{\emph{Phys.Rev.} {\bf
  D88} (2013) 105018}, [\href{https://arxiv.org/abs/1309.5446}{{\tt
  1309.5446}}].

\bibitem{Grahl:2013pba}
M.~Grahl and D.~H. Rischke, \emph{{Functional renormalization group study of
  the two-flavor linear sigma model in the presence of the axial anomaly}},
  \href{http://dx.doi.org/10.1103/PhysRevD.88.056014}{\emph{Phys.Rev.} {\bf
  D88} (2013) 056014}, [\href{https://arxiv.org/abs/1307.2184}{{\tt
  1307.2184}}].

\bibitem{Christ:2003jk}
N.~Christ and X.~Liao, \emph{{Locating the 3-flavor critical point using
  staggered fermions}},
  \href{http://dx.doi.org/10.1016/S0920-5632(03)01600-1}{\emph{Nucl.Phys.Proc.Suppl.}
  {\bf 119} (2003) 514--516}.

\bibitem{deForcrand:2003ut}
P.~de~Forcrand and O.~Philipsen, \emph{{The Phase diagram of N(f) = 3 QCD for
  small baryon densities}},
  \href{http://dx.doi.org/10.1016/S0920-5632(03)02628-8}{\emph{Nucl.Phys.Proc.Suppl.}
  {\bf 129} (2004) 521--523},
  [\href{https://arxiv.org/abs/hep-lat/0309109}{{\tt hep-lat/0309109}}].

\bibitem{deForcrand:2007rq}
P.~de~Forcrand, S.~Kim and O.~Philipsen, \emph{{A QCD chiral critical point at
  small chemical potential: Is it there or not?}}, {\emph{PoS} {\bf LAT2007}
  (2007) 178}, [\href{https://arxiv.org/abs/0711.0262}{{\tt 0711.0262}}].

\bibitem{Smith:2011pm}
D.~Smith and C.~Schmidt, \emph{{On the universal critical behavior in 3-flavor
  QCD}}, {\emph{PoS} {\bf LATTICE2011} (2011) 216},
  [\href{https://arxiv.org/abs/1109.6729}{{\tt 1109.6729}}].

\bibitem{Karsch:2001nf}
F.~Karsch, E.~Laermann and C.~Schmidt, \emph{{The Chiral critical point in
  three-flavor QCD}},
  \href{http://dx.doi.org/10.1016/S0370-2693(01)01114-5}{\emph{Phys.Lett.} {\bf
  B520} (2001) 41--49}, [\href{https://arxiv.org/abs/hep-lat/0107020}{{\tt
  hep-lat/0107020}}].

\bibitem{Schmidt:2002uk}
C.~Schmidt, C.~Allton, S.~Ejiri, S.~Hands, O.~Kaczmarek et~al., \emph{{The
  Quark mass and mu dependence of the QCD chiral critical point}},
  \href{http://dx.doi.org/10.1016/S0920-5632(03)01601-3}{\emph{Nucl.Phys.Proc.Suppl.}
  {\bf 119} (2003) 517--519},
  [\href{https://arxiv.org/abs/hep-lat/0209009}{{\tt hep-lat/0209009}}].

\bibitem{Karsch:2003va}
F.~Karsch, C.~Allton, S.~Ejiri, S.~Hands, O.~Kaczmarek et~al., \emph{{Where is
  the chiral critical point in three flavor QCD?}},
  \href{http://dx.doi.org/10.1016/S0920-5632(03)02659-8}{\emph{Nucl.Phys.Proc.Suppl.}
  {\bf 129} (2004) 614--616},
  [\href{https://arxiv.org/abs/hep-lat/0309116}{{\tt hep-lat/0309116}}].

\bibitem{Bernard:2004je}
{\scshape MILC} collaboration, C.~Bernard et~al., \emph{{QCD thermodynamics
  with three flavors of improved staggered quarks}},
  \href{http://dx.doi.org/10.1103/PhysRevD.71.034504}{\emph{Phys.Rev.} {\bf
  D71} (2005) 034504}, [\href{https://arxiv.org/abs/hep-lat/0405029}{{\tt
  hep-lat/0405029}}].

\bibitem{Cheng:2006aj}
M.~Cheng, N.~Christ, M.~Clark, J.~van~der Heide, C.~Jung et~al., \emph{{Study
  of the finite temperature transition in 3-flavor QCD using the R and RHMC
  algorithms}},
  \href{http://dx.doi.org/10.1103/PhysRevD.75.034506}{\emph{Phys.Rev.} {\bf
  D75} (2007) 034506}, [\href{https://arxiv.org/abs/hep-lat/0612001}{{\tt
  hep-lat/0612001}}].

\bibitem{Endrodi:2007gc}
G.~Endrodi, Z.~Fodor, S.~Katz and K.~Szabo, \emph{{The Nature of the finite
  temperature QCD transition as a function of the quark masses}}, {\emph{PoS}
  {\bf LAT2007} (2007) 182}, [\href{https://arxiv.org/abs/0710.0998}{{\tt
  0710.0998}}].

\bibitem{Jin:2014hea}
X.-Y. Jin, Y.~Kuramashi, Y.~Nakamura, S.~Takeda and A.~Ukawa, \emph{{Critical
  endpoint of the finite temperature phase transition for three flavor QCD}},
  \href{http://dx.doi.org/10.1103/PhysRevD.91.014508}{\emph{Phys.Rev.} {\bf
  D91} (2015) 014508}, [\href{https://arxiv.org/abs/1411.7461}{{\tt
  1411.7461}}].

\bibitem{Varnhorst:2015lea}
L.~Varnhorst, \emph{{The $N_f$=3 critical endpoint with smeared staggered
  quarks}}, {\emph{PoS} {\bf LATTICE2014} (2015) 193}.

\bibitem{Bazavov:2017xul}
A.~Bazavov, H.~T. Ding, P.~Hegde, F.~Karsch, E.~Laermann, S.~Mukherjee et~al.,
  \emph{{Chiral phase structure of three flavor QCD at vanishing baryon number
  density}},  \href{https://arxiv.org/abs/1701.03548}{{\tt 1701.03548}}.

\bibitem{Takeda:2016vfj}
S.~Takeda, X.-Y. Jin, Y.~Kuramashi, Y.~Nakamura and A.~Ukawa, \emph{{Update on
  Nf=3 finite temperature QCD phase structure with Wilson-Clover fermion
  action}},  in \emph{{Proceedings, 34th International Symposium on Lattice
  Field Theory (Lattice 2016): Southampton, UK, July 24-30, 2016}}, 2016.
\newblock \href{https://arxiv.org/abs/1612.05371}{{\tt 1612.05371}}.

\bibitem{Forcrand2016}
P.~de~Forcrand, \emph{Continuum limit and universality of the columbia plot},
  in \emph{{this Proceedings}}, 2016.

\bibitem{Saito:2011fs}
{\scshape WHOT-QCD} collaboration, H.~Saito et~al., \emph{{Phase structure of
  finite temperature QCD in the heavy quark region}},
  \href{http://dx.doi.org/10.1103/PhysRevD.85.079902,
  10.1103/PhysRevD.84.054502}{\emph{Phys.Rev.} {\bf D84} (2011) 054502},
  [\href{https://arxiv.org/abs/1106.0974}{{\tt 1106.0974}}].

\bibitem{Czaban:2016yae}
C.~Czaban and O.~Philipsen, \emph{{The QCD deconfinement critical point for
  $N_\tau=8$ with $N_f=2$ flavours of unimproved Wilson fermions}},
\newblock \href{https://arxiv.org/abs/1609.05745}{{\tt 1609.05745}}.

\bibitem{Iwasaki:1996ya}
Y.~Iwasaki, K.~Kanaya, S.~Kaya and T.~Yoshie, \emph{{Scaling of chiral order
  parameter in two flavor QCD}},
  \href{http://dx.doi.org/10.1103/PhysRevLett.78.179}{\emph{Phys. Rev. Lett.}
  {\bf 78} (1997) 179--182}, [\href{https://arxiv.org/abs/hep-lat/9609022}{{\tt
  hep-lat/9609022}}].

\bibitem{Aoki:1998wg}
{\scshape JLQCD} collaboration, S.~Aoki et~al., \emph{{Scaling study of the two
  flavor chiral phase transition with the Kogut-Susskind quark action in
  lattice QCD}}, \href{http://dx.doi.org/10.1103/PhysRevD.57.3910}{\emph{Phys.
  Rev.} {\bf D57} (1998) 3910--3922},
  [\href{https://arxiv.org/abs/hep-lat/9710048}{{\tt hep-lat/9710048}}].

\bibitem{PhysRevD.63.034502}
{\scshape CP-PACS Collaboration} collaboration, A.~Ali~Khan, S.~Aoki,
  R.~Burkhalter, S.~Ejiri, M.~Fukugita, S.~Hashimoto et~al., \emph{Phase
  structure and critical temperature of two-flavor qcd with a renormalization
  group improved gauge action and clover improved wilson quark action},
  \href{http://dx.doi.org/10.1103/PhysRevD.63.034502}{\emph{Phys. Rev. D} {\bf
  63} (Dec, 2000) 034502}.

\bibitem{Burger:2011zc}
{\scshape tmfT} collaboration, F.~Burger, E.-M. Ilgenfritz, M.~Kirchner, M.~P.
  Lombardo, M.~M{\"u}ller-Preussker, O.~Philipsen et~al., \emph{{Thermal QCD
  transition with two flavors of twisted mass fermions}},
  \href{http://dx.doi.org/10.1103/PhysRevD.87.074508}{\emph{Phys. Rev.} {\bf
  D87} (2013) 074508}, [\href{https://arxiv.org/abs/1102.4530}{{\tt
  1102.4530}}].

\bibitem{Bornyakov:2011yb}
V.~G. Bornyakov, R.~Horsley, Y.~Nakamura, M.~I. Polikarpov, P.~Rakow and
  G.~Schierholz, \emph{{Finite temperature phase transition with two flavors of
  improved Wilson fermions}}, {\emph{PoS} {\bf Lattice2010} (2014) 170},
  [\href{https://arxiv.org/abs/1102.4461}{{\tt 1102.4461}}].

\bibitem{Karsch:1993tv}
F.~Karsch, \emph{{Scaling of pseudocritical couplings in two flavor QCD}},
  \href{http://dx.doi.org/10.1103/PhysRevD.49.3791}{\emph{Phys. Rev.} {\bf D49}
  (1994) 3791--3794}, [\href{https://arxiv.org/abs/hep-lat/9309022}{{\tt
  hep-lat/9309022}}].

\bibitem{Karsch:1994hm}
F.~Karsch and E.~Laermann, \emph{{Susceptibilities, the specific heat and a
  cumulant in two flavor QCD}},
  \href{http://dx.doi.org/10.1103/PhysRevD.50.6954}{\emph{Phys. Rev.} {\bf D50}
  (1994) 6954--6962}, [\href{https://arxiv.org/abs/hep-lat/9406008}{{\tt
  hep-lat/9406008}}].

\bibitem{Laermann:1998gf}
E.~Laermann, \emph{{Chiral transition in 2 flavor staggered QCD}},
  \href{http://dx.doi.org/10.1016/S0920-5632(97)00479-9}{\emph{Nucl. Phys.
  Proc. Suppl.} {\bf 60A} (1998) 180--187}.

\bibitem{Bernard:1996zw}
C.~W. Bernard, T.~Blum, C.~E. Detar, S.~A. Gottlieb, K.~Rummukainen, U.~M.
  Heller et~al., \emph{{Two flavor staggered fermion thermodynamics at N(t) =
  12}}, \href{http://dx.doi.org/10.1103/PhysRevD.54.4585}{\emph{Phys. Rev.}
  {\bf D54} (1996) 4585--4594},
  [\href{https://arxiv.org/abs/hep-lat/9605028}{{\tt hep-lat/9605028}}].

\bibitem{Bernard:1999fv}
C.~W. Bernard, C.~E. Detar, S.~A. Gottlieb, U.~M. Heller, J.~Hetrick,
  K.~Rummukainen et~al., \emph{{Critical behavior in N(t) = 4 staggered fermion
  thermodynamics}},
  \href{http://dx.doi.org/10.1103/PhysRevD.61.054503}{\emph{Phys. Rev.} {\bf
  D61} (2000) 054503}, [\href{https://arxiv.org/abs/hep-lat/9908008}{{\tt
  hep-lat/9908008}}].

\bibitem{DElia:2005bv}
M.~D'Elia, A.~Di~Giacomo and C.~Pica, \emph{{Two flavor QCD and confinement}},
  \href{http://dx.doi.org/10.1103/PhysRevD.72.114510}{\emph{Phys. Rev.} {\bf
  D72} (2005) 114510}, [\href{https://arxiv.org/abs/hep-lat/0503030}{{\tt
  hep-lat/0503030}}].

\bibitem{Mendes:2007ve}
T.~Mendes, \emph{{Universality and Scaling at the chiral transition in
  two-flavor QCD at finite temperature}}, {\emph{PoS} {\bf LAT2007} (2007)
  208}, [\href{https://arxiv.org/abs/0710.0746}{{\tt 0710.0746}}].

\bibitem{LiLattice2016}
S.-T. Li, \emph{Chiral phase transition in (2+1)-flavor qcd on
  $n_\tau=6$lattices},  in \emph{{this proceedings}}, 2016.

\bibitem{Umeda:2016qdo}
T.~Umeda, S.~Ejiri, R.~Iwami, K.~Kanaya, H.~Ohno, A.~Uji et~al., \emph{{O(4)
  scaling analysis in two-flavor QCD at finite temperature and density with
  improved Wilson quarks}},
\newblock \href{https://arxiv.org/abs/1612.09449}{{\tt 1612.09449}}.

\bibitem{Yamada:2016hvz}
N.~Yamada, S.~Ejiri and R.~Iwami, \emph{{Many flavor approach to study the
  nature of chiral phase transition of two-flavor QCD}}, {\emph{PoS} {\bf
  LATTICE2015} (2016) 147}, [\href{https://arxiv.org/abs/1602.04595}{{\tt
  1602.04595}}].

\bibitem{Ejiri:2012rr}
S.~Ejiri and N.~Yamada, \emph{{End Point of a First-Order Phase Transition in
  Many-Flavor Lattice QCD at Finite Temperature and Density}},
  \href{http://dx.doi.org/10.1103/PhysRevLett.110.172001}{\emph{Phys. Rev.
  Lett.} {\bf 110} (2013) 172001}, [\href{https://arxiv.org/abs/1212.5899}{{\tt
  1212.5899}}].

\bibitem{Bonati:2014kpa}
C.~Bonati, P.~de~Forcrand, M.~D'Elia, O.~Philipsen and F.~Sanfilippo,
  \emph{{Chiral phase transition in two-flavor QCD from an imaginary chemical
  potential}},
  \href{http://dx.doi.org/10.1103/PhysRevD.90.074030}{\emph{Phys.Rev.} {\bf
  D90} (2014) 074030}, [\href{https://arxiv.org/abs/1408.5086}{{\tt
  1408.5086}}].

\bibitem{Philipsen:2016hkv}
O.~Philipsen and C.~Pinke, \emph{{The $N_f=2$ QCD chiral phase transition with
  Wilson fermions at zero and imaginary chemical potential}},
  \href{http://dx.doi.org/10.1103/PhysRevD.93.114507}{\emph{Phys. Rev.} {\bf
  D93} (2016) 114507}, [\href{https://arxiv.org/abs/1602.06129}{{\tt
  1602.06129}}].

\bibitem{Bazavov:2012qja}
{\scshape HotQCD} collaboration, A.~Bazavov et~al., \emph{{The chiral
  transition and $U(1)_A$ symmetry restoration from lattice QCD using Domain
  Wall Fermions}},
  \href{http://dx.doi.org/10.1103/PhysRevD.86.094503}{\emph{Phys.Rev.} {\bf
  D86} (2012) 094503}, [\href{https://arxiv.org/abs/1205.3535}{{\tt
  1205.3535}}].

\bibitem{Buchoff:2013nra}
M.~I. Buchoff, M.~Cheng, N.~H. Christ, H.~T. Ding, C.~Jung et~al., \emph{{QCD
  chiral transition, U(1)A symmetry and the dirac spectrum using domain wall
  fermions}},
  \href{http://dx.doi.org/10.1103/PhysRevD.89.054514}{\emph{Phys.Rev.} {\bf
  D89} (2014) 054514}, [\href{https://arxiv.org/abs/1309.4149}{{\tt
  1309.4149}}].

\bibitem{Bhattacharya:2014ara}
T.~Bhattacharya et~al., \emph{{QCD Phase Transition with Chiral Quarks and
  Physical Quark Masses}},
  \href{http://dx.doi.org/10.1103/PhysRevLett.113.082001}{\emph{Phys. Rev.
  Lett.} {\bf 113} (2014) 082001}, [\href{https://arxiv.org/abs/1402.5175}{{\tt
  1402.5175}}].

\bibitem{Chiu:2013wwa}
{\scshape TWQCD} collaboration, T.-W. Chiu, W.-P. Chen, Y.-C. Chen, H.-Y. Chou
  and T.-H. Hsieh, \emph{{Chiral symmetry and axial U(1) symmetry in finite
  temperature QCD with domain-wall fermion}}, {\emph{PoS} {\bf LATTICE2013}
  (2014) 165}, [\href{https://arxiv.org/abs/1311.6220}{{\tt 1311.6220}}].

\bibitem{Cossu:2013uua}
G.~Cossu, S.~Aoki, H.~Fukaya, S.~Hashimoto, T.~Kaneko, H.~Matsufuru et~al.,
  \emph{{Finite temperature study of the axial U(1) symmetry on the lattice
  with overlap fermion formulation}},
  \href{http://dx.doi.org/10.1103/PhysRevD.88.019901,
  10.1103/PhysRevD.87.114514}{\emph{Phys. Rev.} {\bf D87} (2013) 114514},
  [\href{https://arxiv.org/abs/1304.6145}{{\tt 1304.6145}}].

\bibitem{Tomiya:2016jwr}
A.~Tomiya, G.~Cossu, S.~Aoki, H.~Fukaya, S.~Hashimoto, T.~Kaneko et~al.,
  \emph{{Evidence of effective axial U(1) symmetry restoration at high
  temperature QCD}},  \href{https://arxiv.org/abs/1612.01908}{{\tt
  1612.01908}}.

\bibitem{Petreczky:2016vrs}
P.~Petreczky, H.-P. Schadler and S.~Sharma, \emph{{The topological
  susceptibility in finite temperature QCD and axion cosmology}},
  \href{http://dx.doi.org/10.1016/j.physletb.2016.09.063}{\emph{Phys. Lett.}
  {\bf B762} (2016) 498--505}, [\href{https://arxiv.org/abs/1606.03145}{{\tt
  1606.03145}}].

\bibitem{Maezawa2016}
Y.~Maezawa et~al., {\emph{in preparation} }.

\bibitem{Brandt:2016daq}
B.~B. Brandt, A.~Francis, H.~B. Meyer, O.~Philipsen, D.~Robaina and H.~Wittig,
  \emph{{On the strength of the $U_A(1)$ anomaly at the chiral phase transition
  in $N_f=2$ QCD}},
  \href{http://dx.doi.org/10.1007/JHEP12(2016)158}{\emph{JHEP} {\bf 12} (2016)
  158}, [\href{https://arxiv.org/abs/1608.06882}{{\tt 1608.06882}}].

\bibitem{Cheng:2010fe}
M.~Cheng, S.~Datta, A.~Francis, J.~van~der Heide, C.~Jung et~al., \emph{{Meson
  screening masses from lattice QCD with two light and the strange quark}},
  \href{http://dx.doi.org/10.1140/epjc/s10052-011-1564-y}{\emph{Eur.Phys.J.}
  {\bf C71} (2011) 1564}, [\href{https://arxiv.org/abs/1010.1216}{{\tt
  1010.1216}}].

\bibitem{Dick:2015twa}
V.~Dick, F.~Karsch, E.~Laermann, S.~Mukherjee and S.~Sharma, \emph{{Microscopic
  Origin of \boldmath{$U_A(1)$} Symmetry Violation in the High Temperature
  Phase of QCD}},  \href{https://arxiv.org/abs/1502.06190}{{\tt 1502.06190}}.

\bibitem{Bazavov:2016uvm}
A.~Bazavov, N.~Brambilla, H.~T. Ding, P.~Petreczky, H.~P. Schadler, A.~Vairo
  et~al., \emph{{Polyakov loop in 2+1 flavor QCD from low to high
  temperatures}},
  \href{http://dx.doi.org/10.1103/PhysRevD.93.114502}{\emph{Phys. Rev.} {\bf
  D93} (2016) 114502}, [\href{https://arxiv.org/abs/1603.06637}{{\tt
  1603.06637}}].

\bibitem{Cossu:2016scb}
G.~Cossu and S.~Hashimoto, \emph{{Anderson Localization in high temperature
  QCD: background configuration properties and Dirac eigenmodes}},
  \href{http://dx.doi.org/10.1007/JHEP06(2016)056}{\emph{JHEP} {\bf 06} (2016)
  056}, [\href{https://arxiv.org/abs/1604.00768}{{\tt 1604.00768}}].

\bibitem{Bazavov:2013dta}
A.~Bazavov, H.~T. Ding, P.~Hegde, O.~Kaczmarek, F.~Karsch et~al.,
  \emph{{Strangeness at high temperatures: from hadrons to quarks}},
  \href{http://dx.doi.org/10.1103/PhysRevLett.111.082301}{\emph{Phys.Rev.Lett.}
  {\bf 111} (2013) 082301}, [\href{https://arxiv.org/abs/1304.7220}{{\tt
  1304.7220}}].

\bibitem{Bazavov:2014yba}
A.~Bazavov, H.-T. Ding, P.~Hegde, O.~Kaczmarek, F.~Karsch et~al., \emph{{The
  melting and abundance of open charm hadrons}},
  \href{http://dx.doi.org/10.1016/j.physletb.2014.08.034}{\emph{Phys.Lett.}
  {\bf B737} (2014) 210--215}, [\href{https://arxiv.org/abs/1404.4043}{{\tt
  1404.4043}}].

\bibitem{Lo:2013etb}
P.~M. Lo, B.~Friman, O.~Kaczmarek, K.~Redlich and C.~Sasaki, \emph{{Probing
  Deconfinement with Polyakov Loop Susceptibilities}},
  \href{http://dx.doi.org/10.1103/PhysRevD.88.014506}{\emph{Phys. Rev.} {\bf
  D88} (2013) 014506}, [\href{https://arxiv.org/abs/1306.5094}{{\tt
  1306.5094}}].

\bibitem{Doi:2015kje}
T.~M. Doi, K.~Redlich, C.~Sasaki and H.~Suganuma, \emph{{Dirac spectrum
  representation of Polyakov loop fluctuations in lattice QCD}}, {\emph{PoS}
  {\bf LATTICE2015} (2016) 315}, [\href{https://arxiv.org/abs/1511.02039}{{\tt
  1511.02039}}].

\bibitem{Suganuma:2016lnt}
H.~Suganuma, T.~M. Doi, K.~Redlich and C.~Sasaki, \emph{{Interplay between
  Deconfinement and Chiral Properties}},
  \href{https://arxiv.org/abs/1610.02999}{{\tt 1610.02999}}.

\bibitem{Bazavov:2014pvz}
{\scshape HotQCD} collaboration, A.~Bazavov et~al., \emph{{Equation of state in
  ( 2+1 )-flavor QCD}},
  \href{http://dx.doi.org/10.1103/PhysRevD.90.094503}{\emph{Phys.Rev.} {\bf
  D90} (2014) 094503}, [\href{https://arxiv.org/abs/1407.6387}{{\tt
  1407.6387}}].

\bibitem{Borsanyi:2013bia}
S.~Borsanyi, Z.~Fodor, C.~Hoelbling, S.~D. Katz, S.~Krieg et~al., \emph{{Full
  result for the QCD equation of state with 2+1 flavors}},
  \href{http://dx.doi.org/10.1016/j.physletb.2014.01.007}{\emph{Phys.Lett.}
  {\bf B730} (2014) 99--104}, [\href{https://arxiv.org/abs/1309.5258}{{\tt
  1309.5258}}].

\bibitem{Boyd:1996bx}
G.~Boyd, J.~Engels, F.~Karsch, E.~Laermann, C.~Legeland, M.~Lutgemeier et~al.,
  \emph{{Thermodynamics of SU(3) lattice gauge theory}},
  \href{http://dx.doi.org/10.1016/0550-3213(96)00170-8}{\emph{Nucl. Phys.} {\bf
  B469} (1996) 419--444}, [\href{https://arxiv.org/abs/hep-lat/9602007}{{\tt
  hep-lat/9602007}}].

\bibitem{Giusti:2012yj}
L.~Giusti and H.~B. Meyer, \emph{{Implications of Poincare symmetry for thermal
  field theories in finite-volume}},
  \href{http://dx.doi.org/10.1007/JHEP01(2013)140}{\emph{JHEP} {\bf 1301}
  (2013) 140}, [\href{https://arxiv.org/abs/1211.6669}{{\tt 1211.6669}}].

\bibitem{Giusti:2014ila}
L.~Giusti and M.~Pepe, \emph{{Equation of state of a relativistic theory from a
  moving frame}},
  \href{http://dx.doi.org/10.1103/PhysRevLett.113.031601}{\emph{Phys. Rev.
  Lett.} {\bf 113} (2014) 031601}, [\href{https://arxiv.org/abs/1403.0360}{{\tt
  1403.0360}}].

\bibitem{Giusti:2016iqr}
L.~Giusti and M.~Pepe, \emph{{Equation of state of the SU($3$) Yang-Mills
  theory: a precise determination from a moving frame}},
  \href{https://arxiv.org/abs/1612.00265}{{\tt 1612.00265}}.

\bibitem{Giusti:2016wsf}
L.~Giusti and M.~Pepe, \emph{{Thermodynamics of strongly interacting plasma
  with high accuracy}},
\newblock \href{https://arxiv.org/abs/1612.02337}{{\tt 1612.02337}}.

\bibitem{Luscher:2010iy}
M.~L{\"u}scher, \emph{{Properties and uses of the Wilson flow in lattice QCD}},
  \href{http://dx.doi.org/10.1007/JHEP08(2010)071,
  10.1007/JHEP03(2014)092}{\emph{JHEP} {\bf 08} (2010) 071},
  [\href{https://arxiv.org/abs/1006.4518}{{\tt 1006.4518}}].

\bibitem{Luscher:2011bx}
M.~Luscher and P.~Weisz, \emph{{Perturbative analysis of the gradient flow in
  non-abelian gauge theories}},
  \href{http://dx.doi.org/10.1007/JHEP02(2011)051}{\emph{JHEP} {\bf 02} (2011)
  051}, [\href{https://arxiv.org/abs/1101.0963}{{\tt 1101.0963}}].

\bibitem{Luscher:2013vga}
M.~L{\"u}scher, \emph{{Future applications of the Yang-Mills gradient flow in
  lattice QCD}}, {\emph{PoS} {\bf LATTICE2013} (2014) 016},
  [\href{https://arxiv.org/abs/1308.5598}{{\tt 1308.5598}}].

\bibitem{Suzuki:2013gza}
H.~Suzuki, \emph{{Energy--momentum tensor from the Yang--Mills gradient flow}},
  \href{http://dx.doi.org/10.1093/ptep/ptt059, 10.1093/ptep/ptv094}{\emph{PTEP}
  {\bf 2013} (2013) 083B03}, [\href{https://arxiv.org/abs/1304.0533}{{\tt
  1304.0533}}].

\bibitem{Asakawa:2013laa}
{\scshape FlowQCD} collaboration, M.~Asakawa, T.~Hatsuda, E.~Itou, M.~Kitazawa
  and H.~Suzuki, \emph{{Thermodynamics of SU(3) gauge theory from gradient flow
  on the lattice}}, \href{http://dx.doi.org/10.1103/PhysRevD.90.011501,
  10.1103/PhysRevD.92.059902}{\emph{Phys. Rev.} {\bf D90} (2014) 011501},
  [\href{https://arxiv.org/abs/1312.7492}{{\tt 1312.7492}}].

\bibitem{Kitazawa:2016dsl}
M.~Kitazawa, T.~Iritani, M.~Asakawa, T.~Hatsuda and H.~Suzuki, \emph{{Equation
  of State for SU(3) Gauge Theory via the Energy-Momentum Tensor under Gradient
  Flow}}, \href{http://dx.doi.org/10.1103/PhysRevD.94.114512}{\emph{Phys. Rev.}
  {\bf D94} (2016) 114512}, [\href{https://arxiv.org/abs/1610.07810}{{\tt
  1610.07810}}].

\bibitem{Makino:2014taa}
H.~Makino and H.~Suzuki, \emph{{Lattice energy--momentum tensor from the
  Yang--Mills gradient flow---inclusion of fermion fields}},
  \href{http://dx.doi.org/10.1093/ptep/ptu070, 10.1093/ptep/ptv095}{\emph{PTEP}
  {\bf 2014} (2014) 063B02}, [\href{https://arxiv.org/abs/1403.4772}{{\tt
  1403.4772}}].

\bibitem{Taniguchi:2016ofw}
Y.~Taniguchi, S.~Ejiri, R.~Iwami, K.~Kanaya, M.~Kitazawa, H.~Suzuki et~al.,
  \emph{{Nf=2+1 QCD thermodynamics from gradient flow}},
  \href{https://arxiv.org/abs/1609.01417}{{\tt 1609.01417}}.

\bibitem{Kanaya:2016rkt}
K.~Kanaya, S.~Ejiri, R.~Iwami, M.~Kitazawa, H.~Suzuki, Y.~Taniguchi et~al.,
  \emph{{Equation of state in (2+1)-flavor QCD with gradient flow}},
\newblock \href{https://arxiv.org/abs/1610.09518}{{\tt 1610.09518}}.

\bibitem{Suzuki:2016ytc}
H.~Suzuki, \emph{{Energy--momentum tensor on the lattice: recent
  developments}},  in \emph{{Proceedings, 34th International Symposium on
  Lattice Field Theory (Lattice 2016): Southampton, UK, July 24-30, 2016}},
  2016.
\newblock \href{https://arxiv.org/abs/1612.00210}{{\tt 1612.00210}}.

\bibitem{Bazavov:2012kf}
{\scshape MILC} collaboration, A.~Bazavov et~al., \emph{{Towards a QCD equation
  of state with 2 + 1 + 1 flavors using the HISQ action}}, {\emph{PoS} {\bf
  LATTICE2012} (2012) 071}.

\bibitem{Bazavov:2013pra}
{\scshape MILC} collaboration, A.~Bazavov et~al., \emph{{Update on the 2+1+1
  flavor QCD equation of state with HISQ}}, {\emph{PoS} {\bf LATTICE2013}
  (2014) 154}, [\href{https://arxiv.org/abs/1312.5011}{{\tt 1312.5011}}].

\bibitem{Ratti:2013uta}
C.~Ratti, S.~Borsanyi, G.~Endrodi, Z.~Fodor, S.~D. Katz, S.~Krieg et~al.,
  \emph{{Lattice QCD thermodynamics in the presence of the charm quark}},
  \href{http://dx.doi.org/10.1016/j.nuclphysa.2013.02.153}{\emph{Nucl. Phys.}
  {\bf A904-905} (2013) 869c--872c}.

\bibitem{Burger:2015xda}
F.~Burger, E.-M. Ilgenfritz, M.~P. Lombardo, M.~Muller-Preussker and A.~Trunin,
  \emph{{Towards the quark--gluon plasma Equation of State with dynamical
  strange and charm quarks}},
  \href{http://dx.doi.org/10.1088/1742-6596/668/1/012092}{\emph{J. Phys. Conf.
  Ser.} {\bf 668} (2016) 012092}, [\href{https://arxiv.org/abs/1510.02262}{{\tt
  1510.02262}}].

\bibitem{Borsanyi:2016ksw}
S.~Borsanyi et~al., \emph{{Calculation of the axion mass based on
  high-temperature lattice quantum chromodynamics}},
  \href{http://dx.doi.org/10.1038/nature20115}{\emph{Nature} {\bf 539} (2016)
  69--71}, [\href{https://arxiv.org/abs/1606.07494}{{\tt 1606.07494}}].

\bibitem{Kajantie:2002wa}
K.~Kajantie, M.~Laine, K.~Rummukainen and Y.~Schroder, \emph{{The Pressure of
  hot QCD up to g6 ln(1/g)}},
  \href{http://dx.doi.org/10.1103/PhysRevD.67.105008}{\emph{Phys. Rev.} {\bf
  D67} (2003) 105008}, [\href{https://arxiv.org/abs/hep-ph/0211321}{{\tt
  hep-ph/0211321}}].

\bibitem{Hietanen:2008tv}
A.~Hietanen, K.~Kajantie, M.~Laine, K.~Rummukainen and Y.~Schroder,
  \emph{{Three-dimensional physics and the pressure of hot QCD}},
  \href{http://dx.doi.org/10.1103/PhysRevD.79.045018}{\emph{Phys.Rev.} {\bf
  D79} (2009) 045018}, [\href{https://arxiv.org/abs/0811.4664}{{\tt
  0811.4664}}].

\bibitem{Gross:1980br}
D.~J. Gross, R.~D. Pisarski and L.~G. Yaffe, \emph{{QCD and Instantons at
  Finite Temperature}},
  \href{http://dx.doi.org/10.1103/RevModPhys.53.43}{\emph{Rev.Mod.Phys.} {\bf
  53} (1981) 43}.

\bibitem{Berkowitz:2015aua}
E.~Berkowitz, M.~I. Buchoff and E.~Rinaldi, \emph{{Lattice QCD input for axion
  cosmology}}, \href{http://dx.doi.org/10.1103/PhysRevD.92.034507}{\emph{Phys.
  Rev.} {\bf D92} (2015) 034507}, [\href{https://arxiv.org/abs/1505.07455}{{\tt
  1505.07455}}].

\bibitem{Kitano:2015fla}
R.~Kitano and N.~Yamada, \emph{{Topology in QCD and the axion abundance}},
  \href{http://dx.doi.org/10.1007/JHEP10(2015)136}{\emph{JHEP} {\bf 10} (2015)
  136}, [\href{https://arxiv.org/abs/1506.00370}{{\tt 1506.00370}}].

\bibitem{Borsanyi:2015cka}
S.~Borsanyi, M.~Dierigl, Z.~Fodor, S.~D. Katz, S.~W. Mages, D.~Nogradi et~al.,
  \emph{{Axion cosmology, lattice QCD and the dilute instanton gas}},
  \href{http://dx.doi.org/10.1016/j.physletb.2015.11.020}{\emph{Phys. Lett.}
  {\bf B752} (2016) 175--181}, [\href{https://arxiv.org/abs/1508.06917}{{\tt
  1508.06917}}].

\bibitem{Bonati:2015vqz}
C.~Bonati, M.~D'Elia, M.~Mariti, G.~Martinelli, M.~Mesiti, F.~Negro et~al.,
  \emph{{Axion phenomenology and $\theta$-dependence from $N_f = 2+1$ lattice
  QCD}}, \href{http://dx.doi.org/10.1007/JHEP03(2016)155}{\emph{JHEP} {\bf 03}
  (2016) 155}, [\href{https://arxiv.org/abs/1512.06746}{{\tt 1512.06746}}].

\bibitem{Taniguchi:2016tjc}
Y.~Taniguchi, K.~Kanaya, H.~Suzuki and T.~Umeda, \emph{{Topological
  susceptibility in finite temperature (2+1)-flavor QCD using gradient flow}},
  \href{https://arxiv.org/abs/1611.02411}{{\tt 1611.02411}}.

\bibitem{Marsh:2015xka}
D.~J.~E. Marsh, \emph{{Axion Cosmology}},
  \href{http://dx.doi.org/10.1016/j.physrep.2016.06.005}{\emph{Phys. Rept.}
  {\bf 643} (2016) 1--79}, [\href{https://arxiv.org/abs/1510.07633}{{\tt
  1510.07633}}].

\bibitem{Vicari:2008jw}
E.~Vicari and H.~Panagopoulos, \emph{{Theta dependence of SU(N) gauge theories
  in the presence of a topological term}},
  \href{http://dx.doi.org/10.1016/j.physrep.2008.10.001}{\emph{Phys. Rept.}
  {\bf 470} (2009) 93--150}, [\href{https://arxiv.org/abs/0803.1593}{{\tt
  0803.1593}}].

\bibitem{Ringwald:1999ze}
A.~Ringwald and F.~Schrempp, \emph{{Confronting instanton perturbation theory
  with QCD lattice results}},
  \href{http://dx.doi.org/10.1016/S0370-2693(99)00682-6}{\emph{Phys. Lett.}
  {\bf B459} (1999) 249--258},
  [\href{https://arxiv.org/abs/hep-lat/9903039}{{\tt hep-lat/9903039}}].

\bibitem{Bautista:2015yza}
I.~Bautista, W.~Bietenholz, A.~Dromard, U.~Gerber, L.~Gonglach, C.~P. Hofmann
  et~al., \emph{{Measuring the Topological Susceptibility in a Fixed Sector}},
  \href{http://dx.doi.org/10.1103/PhysRevD.92.114510}{\emph{Phys. Rev.} {\bf
  D92} (2015) 114510}, [\href{https://arxiv.org/abs/1503.06853}{{\tt
  1503.06853}}].

\bibitem{Bietenholz:2015rsa}
W.~Bietenholz, P.~de~Forcrand and U.~Gerber, \emph{{Topological Susceptibility
  from Slabs}}, \href{http://dx.doi.org/10.1007/JHEP12(2015)070}{\emph{JHEP}
  {\bf 12} (2015) 070}, [\href{https://arxiv.org/abs/1509.06433}{{\tt
  1509.06433}}].

\bibitem{Frison:2016vuc}
J.~Frison, R.~Kitano, H.~Matsufuru, S.~Mori and N.~Yamada, \emph{{Topological
  susceptibility at high temperature on the lattice}},
  \href{http://dx.doi.org/10.1007/JHEP09(2016)021}{\emph{JHEP} {\bf 09} (2016)
  021}, [\href{https://arxiv.org/abs/1606.07175}{{\tt 1606.07175}}].

\bibitem{Sharma:2016cmz}
S.~Sharma, V.~Dick, F.~Karsch, E.~Laermann and S.~Mukherjee, \emph{{The
  topological structures in strongly coupled QGP with chiral fermions on the
  lattice}},
  \href{http://dx.doi.org/10.1016/j.nuclphysa.2016.02.013}{\emph{Nucl. Phys.}
  {\bf A956} (2016) 793--796}, [\href{https://arxiv.org/abs/1602.02197}{{\tt
  1602.02197}}].

\bibitem{Cleymans:2005xv}
J.~Cleymans, H.~Oeschler, K.~Redlich and S.~Wheaton, \emph{{Comparison of
  chemical freeze-out criteria in heavy-ion collisions}},
  \href{http://dx.doi.org/10.1103/PhysRevC.73.034905}{\emph{Phys. Rev.} {\bf
  C73} (2006) 034905}, [\href{https://arxiv.org/abs/hep-ph/0511094}{{\tt
  hep-ph/0511094}}].

\bibitem{Andronic:2008gu}
A.~Andronic, P.~Braun-Munzinger and J.~Stachel, \emph{{Thermal hadron
  production in relativistic nuclear collisions: The Hadron mass spectrum, the
  horn, and the QCD phase transition}},
  \href{http://dx.doi.org/10.1016/j.physletb.2009.02.014,
  10.1016/j.physletb.2009.06.021}{\emph{Phys. Lett.} {\bf B673} (2009)
  142--145}, [\href{https://arxiv.org/abs/0812.1186}{{\tt 0812.1186}}].

\bibitem{Allton:2002zi}
C.~Allton, S.~Ejiri, S.~Hands, O.~Kaczmarek, F.~Karsch et~al., \emph{{The QCD
  thermal phase transition in the presence of a small chemical potential}},
  \href{http://dx.doi.org/10.1103/PhysRevD.66.074507}{\emph{Phys.Rev.} {\bf
  D66} (2002) 074507}, [\href{https://arxiv.org/abs/hep-lat/0204010}{{\tt
  hep-lat/0204010}}].

\bibitem{Gavai:2003mf}
R.~V. Gavai and S.~Gupta, \emph{{Pressure and nonlinear susceptibilities in QCD
  at finite chemical potentials}},
  \href{http://dx.doi.org/10.1103/PhysRevD.68.034506}{\emph{Phys.Rev.} {\bf
  D68} (2003) 034506}, [\href{https://arxiv.org/abs/hep-lat/0303013}{{\tt
  hep-lat/0303013}}].

\bibitem{deForcrand:2002hgr}
P.~de~Forcrand and O.~Philipsen, \emph{{The QCD phase diagram for small
  densities from imaginary chemical potential}},
  \href{http://dx.doi.org/10.1016/S0550-3213(02)00626-0}{\emph{Nucl. Phys.}
  {\bf B642} (2002) 290--306},
  [\href{https://arxiv.org/abs/hep-lat/0205016}{{\tt hep-lat/0205016}}].

\bibitem{DElia:2002tig}
M.~D'Elia and M.-P. Lombardo, \emph{{Finite density QCD via imaginary chemical
  potential}}, \href{http://dx.doi.org/10.1103/PhysRevD.67.014505}{\emph{Phys.
  Rev.} {\bf D67} (2003) 014505},
  [\href{https://arxiv.org/abs/hep-lat/0209146}{{\tt hep-lat/0209146}}].

\bibitem{Bazavov:2017dus}
A.~Bazavov et~al., \emph{{The QCD Equation of State to $\mathcal{O}(\mu_B^6)$
  from Lattice QCD}},  \href{https://arxiv.org/abs/1701.04325}{{\tt
  1701.04325}}.

\bibitem{Gunther:2016vcp}
J.~Gunther, R.~Bellwied, S.~Borsanyi, Z.~Fodor, S.~D. Katz, A.~Pasztor et~al.,
  \emph{{The QCD equation of state at finite density from analytical
  continuation}},  \href{https://arxiv.org/abs/1607.02493}{{\tt 1607.02493}}.

\bibitem{Gavai:2011uk}
R.~V. Gavai and S.~Sharma, \emph{{A faster method of computation of lattice
  quark number susceptibilities}},
  \href{http://dx.doi.org/10.1103/PhysRevD.85.054508}{\emph{Phys. Rev.} {\bf
  D85} (2012) 054508}, [\href{https://arxiv.org/abs/1112.5428}{{\tt
  1112.5428}}].

\bibitem{Gavai:2014lia}
R.~V. Gavai and S.~Sharma, \emph{{Divergences in the quark number
  susceptibility: The origin and a cure}},
  \href{http://dx.doi.org/10.1016/j.physletb.2015.07.036}{\emph{Phys. Lett.}
  {\bf B749} (2015) 8--13}, [\href{https://arxiv.org/abs/1406.0474}{{\tt
  1406.0474}}].

\bibitem{Bellwied:2015rza}
R.~Bellwied, S.~Borsanyi, Z.~Fodor, J.~G{\"u}nther, S.~D. Katz, C.~Ratti
  et~al., \emph{{The QCD phase diagram from analytic continuation}},
  \href{http://dx.doi.org/10.1016/j.physletb.2015.11.011}{\emph{Phys. Lett.}
  {\bf B751} (2015) 559--564}, [\href{https://arxiv.org/abs/1507.07510}{{\tt
  1507.07510}}].

\bibitem{Luo:2017faz}
X.~Luo and N.~Xu, \emph{{Search for the QCD Critical Point with Fluctuations of
  Conserved Quantities in Relativistic Heavy-Ion Collisions at RHIC : An
  Overview}},  \href{https://arxiv.org/abs/1701.02105}{{\tt 1701.02105}}.

\bibitem{Borsanyi:2011sw}
S.~Borsanyi, Z.~Fodor, S.~D. Katz, S.~Krieg, C.~Ratti et~al.,
  \emph{{Fluctuations of conserved charges at finite temperature from lattice
  QCD}}, \href{http://dx.doi.org/10.1007/JHEP01(2012)138}{\emph{JHEP} {\bf
  1201} (2012) 138}, [\href{https://arxiv.org/abs/1112.4416}{{\tt 1112.4416}}].

\bibitem{Bazavov:2012jq}
{\scshape HotQCD} collaboration, A.~Bazavov et~al., \emph{{Fluctuations and
  Correlations of net baryon number, electric charge, and strangeness: A
  comparison of lattice QCD results with the hadron resonance gas model}},
  \href{http://dx.doi.org/10.1103/PhysRevD.86.034509}{\emph{Phys.Rev.} {\bf
  D86} (2012) 034509}, [\href{https://arxiv.org/abs/1203.0784}{{\tt
  1203.0784}}].

\bibitem{Bazavov:2013uja}
A.~Bazavov, H.-T. Ding, P.~Hegde, F.~Karsch, C.~Miao et~al., \emph{{Quark
  number susceptibilities at high temperatures}},
  \href{http://dx.doi.org/10.1103/PhysRevD.88.094021}{\emph{Phys.Rev.} {\bf
  D88} (2013) 094021}, [\href{https://arxiv.org/abs/1309.2317}{{\tt
  1309.2317}}].

\bibitem{Ding:2015fca}
H.~T. Ding, S.~Mukherjee, H.~Ohno, P.~Petreczky and H.~P. Schadler,
  \emph{{Diagonal and off-diagonal quark number susceptibilities at high
  temperatures}},
  \href{http://dx.doi.org/10.1103/PhysRevD.92.074043}{\emph{Phys. Rev.} {\bf
  D92} (2015) 074043}, [\href{https://arxiv.org/abs/1507.06637}{{\tt
  1507.06637}}].

\bibitem{Bellwied:2015lba}
R.~Bellwied, S.~Borsanyi, Z.~Fodor, S.~D. Katz, A.~Pasztor, C.~Ratti et~al.,
  \emph{{Fluctuations and correlations in high temperature QCD}},
  \href{http://dx.doi.org/10.1103/PhysRevD.92.114505}{\emph{Phys. Rev.} {\bf
  D92} (2015) 114505}, [\href{https://arxiv.org/abs/1507.04627}{{\tt
  1507.04627}}].

\bibitem{Friman:2011pf}
B.~Friman, F.~Karsch, K.~Redlich and V.~Skokov, \emph{{Fluctuations as probe of
  the QCD phase transition and freeze-out in heavy ion collisions at LHC and
  RHIC}},
  \href{http://dx.doi.org/10.1140/epjc/s10052-011-1694-2}{\emph{Eur.Phys.J.}
  {\bf C71} (2011) 1694}, [\href{https://arxiv.org/abs/1103.3511}{{\tt
  1103.3511}}].

\bibitem{Kaczmarek:2011zz}
O.~Kaczmarek, F.~Karsch, E.~Laermann, C.~Miao, S.~Mukherjee et~al.,
  \emph{{Phase boundary for the chiral transition in (2+1) -flavor QCD at small
  values of the chemical potential}},
  \href{http://dx.doi.org/10.1103/PhysRevD.83.014504}{\emph{Phys.Rev.} {\bf
  D83} (2011) 014504}, [\href{https://arxiv.org/abs/1011.3130}{{\tt
  1011.3130}}].

\bibitem{Endrodi:2011gv}
G.~Endrodi, Z.~Fodor, S.~Katz and K.~Szabo, \emph{{The QCD phase diagram at
  nonzero quark density}},
  \href{http://dx.doi.org/10.1007/JHEP04(2011)001}{\emph{JHEP} {\bf 1104}
  (2011) 001}, [\href{https://arxiv.org/abs/1102.1356}{{\tt 1102.1356}}].

\bibitem{Bonati:2014rfa}
C.~Bonati, M.~D'Elia, M.~Mariti, M.~Mesiti, F.~Negro et~al., \emph{{Curvature
  of the chiral pseudocritical line in QCD}},
  \href{http://dx.doi.org/10.1103/PhysRevD.90.114025}{\emph{Phys.Rev.} {\bf
  D90} (2014) 114025}, [\href{https://arxiv.org/abs/1410.5758}{{\tt
  1410.5758}}].

\bibitem{Cea:2015cya}
P.~Cea, L.~Cosmai and A.~Papa, \emph{{Critical line of 2+1 flavor QCD: Toward
  the continuum limit}},
  \href{http://dx.doi.org/10.1103/PhysRevD.93.014507}{\emph{Phys. Rev.} {\bf
  D93} (2016) 014507}, [\href{https://arxiv.org/abs/1508.07599}{{\tt
  1508.07599}}].

\bibitem{Bazavov:2012vg}
A.~Bazavov, H.~Ding, P.~Hegde, O.~Kaczmarek, F.~Karsch et~al.,
  \emph{{Freeze-out Conditions in Heavy Ion Collisions from QCD
  Thermodynamics}},
  \href{http://dx.doi.org/10.1103/PhysRevLett.109.192302}{\emph{Phys.Rev.Lett.}
  {\bf 109} (2012) 192302}, [\href{https://arxiv.org/abs/1208.1220}{{\tt
  1208.1220}}].

\bibitem{Bazavov:2014xya}
A.~Bazavov, H.~T. Ding, P.~Hegde, O.~Kaczmarek, F.~Karsch et~al.,
  \emph{{Additional Strange Hadrons from QCD Thermodynamics and Strangeness
  Freezeout in Heavy Ion Collisions}},
  \href{http://dx.doi.org/10.1103/PhysRevLett.113.072001}{\emph{Phys.Rev.Lett.}
  {\bf 113} (2014) 072001}, [\href{https://arxiv.org/abs/1404.6511}{{\tt
  1404.6511}}].

\bibitem{Bazavov:2015zja}
A.~Bazavov et~al., \emph{{Curvature of the freeze-out line in heavy ion
  collisions}}, \href{http://dx.doi.org/10.1103/PhysRevD.93.014512}{\emph{Phys.
  Rev.} {\bf D93} (2016) 014512}, [\href{https://arxiv.org/abs/1509.05786}{{\tt
  1509.05786}}].

\bibitem{Borsanyi:2013hza}
S.~Borsanyi, Z.~Fodor, S.~Katz, S.~Krieg, C.~Ratti et~al., \emph{{Freeze-out
  parameters: lattice meets experiment}},
  \href{http://dx.doi.org/10.1103/PhysRevLett.111.062005}{\emph{Phys.Rev.Lett.}
  {\bf 111} (2013) 062005}, [\href{https://arxiv.org/abs/1305.5161}{{\tt
  1305.5161}}].

\bibitem{Borsanyi:2014ewa}
S.~Borsanyi, Z.~Fodor, S.~Katz, S.~Krieg, C.~Ratti et~al., \emph{{Freeze-out
  parameters from electric charge and baryon number fluctuations: is there
  consistency?}},
  \href{http://dx.doi.org/10.1103/PhysRevLett.113.052301}{\emph{Phys.Rev.Lett.}
  {\bf 113} (2014) 052301}, [\href{https://arxiv.org/abs/1403.4576}{{\tt
  1403.4576}}].

\bibitem{Bellwied:2016cpq}
R.~Bellwied, S.~Borsanyi, Z.~Fodor, J.~Gunther, S.~D. Katz, A.~Pasztor et~al.,
  \emph{{Towards the QCD phase diagram from analytical continuation}},
  \href{http://dx.doi.org/10.1016/j.nuclphysa.2016.02.010}{\emph{Nucl. Phys.}
  {\bf A956} (2016) 797--800}, [\href{https://arxiv.org/abs/1601.00466}{{\tt
  1601.00466}}].

\bibitem{Cleymans:1999st}
J.~Cleymans and K.~Redlich, \emph{{Chemical and thermal freezeout parameters
  from 1-A/GeV to 200-A/GeV}},
  \href{http://dx.doi.org/10.1103/PhysRevC.60.054908}{\emph{Phys. Rev.} {\bf
  C60} (1999) 054908}, [\href{https://arxiv.org/abs/nucl-th/9903063}{{\tt
  nucl-th/9903063}}].

\bibitem{Das:2014qca}
{\scshape STAR} collaboration, S.~Das, \emph{{Identified particle production
  and freeze-out properties in heavy-ion collisions at RHIC Beam Energy Scan
  program}},  \href{https://arxiv.org/abs/1412.0499}{{\tt 1412.0499}}.

\bibitem{Floris:2014pta}
M.~Floris, \emph{{Hadron yields and the phase diagram of strongly interacting
  matter}},
  \href{http://dx.doi.org/10.1016/j.nuclphysa.2014.09.002}{\emph{Nucl. Phys.}
  {\bf A931} (2014) 103--112}, [\href{https://arxiv.org/abs/1408.6403}{{\tt
  1408.6403}}].

\bibitem{Becattini:2016xct}
F.~Becattini, J.~Steinheimer, R.~Stock and M.~Bleicher, \emph{{Hadronization
  conditions in relativistic nuclear collisions and the QCD pseudo-critical
  line}}, \href{http://dx.doi.org/10.1016/j.physletb.2016.11.033}{\emph{Phys.
  Lett.} {\bf B764} (2017) 241--246},
  [\href{https://arxiv.org/abs/1605.09694}{{\tt 1605.09694}}].

\bibitem{DElia:2016jqh}
M.~D'Elia, G.~Gagliardi and F.~Sanfilippo, \emph{{Higher order quark number
  fluctuations via imaginary chemical potentials in $N_f=2+1$ QCD}},
  \href{https://arxiv.org/abs/1611.08285}{{\tt 1611.08285}}.

\bibitem{Fodor:2004nz}
Z.~Fodor and S.~Katz, \emph{{Critical point of QCD at finite T and mu, lattice
  results for physical quark masses}},
  \href{http://dx.doi.org/10.1088/1126-6708/2004/04/050}{\emph{JHEP} {\bf 0404}
  (2004) 050}, [\href{https://arxiv.org/abs/hep-lat/0402006}{{\tt
  hep-lat/0402006}}].

\bibitem{Datta:2016ukp}
S.~Datta, R.~V. Gavai and S.~Gupta, \emph{{Quark number susceptibilities and
  equation of state at finite chemical potential in staggered QCD with Nt=8}},
  \href{https://arxiv.org/abs/1612.06673}{{\tt 1612.06673}}.

\bibitem{Cristoforetti:2012su}
{\scshape AuroraScience} collaboration, M.~Cristoforetti, F.~Di~Renzo and
  L.~Scorzato, \emph{{New approach to the sign problem in quantum field
  theories: High density QCD on a Lefschetz thimble}},
  \href{http://dx.doi.org/10.1103/PhysRevD.86.074506}{\emph{Phys.Rev.} {\bf
  D86} (2012) 074506}, [\href{https://arxiv.org/abs/1205.3996}{{\tt
  1205.3996}}].

\bibitem{Aarts:2009uq}
G.~Aarts, E.~Seiler and I.-O. Stamatescu, \emph{{The Complex Langevin method:
  When can it be trusted?}},
  \href{http://dx.doi.org/10.1103/PhysRevD.81.054508}{\emph{Phys. Rev.} {\bf
  D81} (2010) 054508}, [\href{https://arxiv.org/abs/0912.3360}{{\tt
  0912.3360}}].

\bibitem{Aarts:2017vrv}
G.~Aarts, E.~Seiler, D.~Sexty and I.-O. Stamatescu, \emph{{Complex Langevin
  dynamics and zeroes of the fermion determinant}},
  \href{https://arxiv.org/abs/1701.02322}{{\tt 1701.02322}}.

\bibitem{Langfeld:2012ah}
K.~Langfeld, B.~Lucini and A.~Rago, \emph{{The density of states in gauge
  theories}},
  \href{http://dx.doi.org/10.1103/PhysRevLett.109.111601}{\emph{Phys. Rev.
  Lett.} {\bf 109} (2012) 111601}, [\href{https://arxiv.org/abs/1204.3243}{{\tt
  1204.3243}}].

\bibitem{Langfeld:2016kty}
K.~Langfeld, \emph{{Density-of-states}},  in \emph{{Proceedings, 34th
  International Symposium on Lattice Field Theory (Lattice 2016): Southampton,
  UK, July 24-30, 2016}}, 2016.
\newblock \href{https://arxiv.org/abs/1610.09856}{{\tt 1610.09856}}.

\bibitem{Nakamura:2015jra}
A.~Nakamura, S.~Oka and Y.~Taniguchi, \emph{{QCD phase transition at real
  chemical potential with canonical approach}},
  \href{http://dx.doi.org/10.1007/JHEP02(2016)054}{\emph{JHEP} {\bf 02} (2016)
  054}, [\href{https://arxiv.org/abs/1504.04471}{{\tt 1504.04471}}].

\bibitem{Bornyakov:2016wld}
V.~G. Bornyakov, D.~L. Boyda, V.~A. Goy, A.~V. Molochkov, A.~Nakamura, A.~A.
  Nikolaev et~al., \emph{{New approach to canonical partition functions
  computation in $N_f=2$ lattice QCD at finite baryon density}},
  \href{https://arxiv.org/abs/1611.04229}{{\tt 1611.04229}}.

\bibitem{Brandt:2016zdy}
B.~B. Brandt and G.~Endrodi, \emph{{QCD phase diagram with isospin chemical
  potential}},  in \emph{{Proceedings, 34th International Symposium on Lattice
  Field Theory (Lattice 2016): Southampton, UK, July 24-30, 2016}}, 2016.
\newblock \href{https://arxiv.org/abs/1611.06758}{{\tt 1611.06758}}.

\bibitem{Bonati:2016pwz}
C.~Bonati, M.~D'Elia, M.~Mariti, M.~Mesiti, F.~Negro and F.~Sanfilippo,
  \emph{{Roberge-Weiss endpoint at the physical point of $N_f = 2+1$ QCD}},
  \href{http://dx.doi.org/10.1103/PhysRevD.93.074504}{\emph{Phys. Rev.} {\bf
  D93} (2016) 074504}, [\href{https://arxiv.org/abs/1602.01426}{{\tt
  1602.01426}}].

\bibitem{Cuteri:2015qkq}
C.~Czaban, F.~Cuteri, O.~Philipsen, C.~Pinke and A.~Sciarra,
  \emph{{Roberge-Weiss transition in $N_f=2$ QCD with Wilson fermions and
  $N_\tau=6$}}, \href{http://dx.doi.org/10.1103/PhysRevD.93.054507}{\emph{Phys.
  Rev.} {\bf D93} (2016) 054507}, [\href{https://arxiv.org/abs/1512.07180}{{\tt
  1512.07180}}].

\bibitem{Philipsen:2016swy}
O.~Philipsen and A.~Sciarra, \emph{{Roberge-Weiss transition in $N_f=2$ QCD
  with staggered fermions and $N_\tau=6$}},
\newblock \href{https://arxiv.org/abs/1610.09979}{{\tt 1610.09979}}.

\end{thebibliography}

\providecommand{\href}[2]{#2}\begingroup\raggedright\endgroup

\end{document}